\documentclass[%
 aip,
 jcp,%
 reprint,%
 superscriptaddress,
 citeautoscript,
]{revtex4-2}
\usepackage{xcolor}
\usepackage{amsmath,amssymb,amsfonts}
\usepackage[normalem]{ulem}
 \usepackage{physics}
 \usepackage{graphicx}
\usepackage{dcolumn}
\usepackage{bm}
\usepackage{hyperref}

\newif\ifshowedits
\showeditstrue     

\ifshowedits
  
\else
  
\fi

\newif\iferbedit
\erbedittrue   
\iferbedit
  
  \newcommand{\erbdel}[1]{\textcolor{red}{\sout{#1}}}
  
\else
  
  \newcommand{\erbdel}[1]{}
  
\fi

\begin{document}



\title{Detection Defines Dephasing in Two-Dimensional Electronic Spectroscopy of Materials: Coherent Field Emission versus Incoherent Population Observables}

\author{Sim\'on~Paiva-Ortega}
\affiliation{Institut Courtois, Universit\'e de Montr\'eal, 1375 Avenue Th\'er\`ese-Lavoie-Roux, Montr\'eal H2V~0B3, Qu\'ebec, Canada}%
\affiliation{D\'epartement de physique, Universit\'e de Montr\'eal, 1375 Avenue Th\'er\`ese-Lavoie-Roux, Montr\'eal H2V~0B3, Qu\'ebec, Canada}%

\author{Hao~Li}
\email{hao.li.3@umontreal.ca}
\affiliation{Institut Courtois, Universit\'e de Montr\'eal, 1375 Avenue Th\'er\`ese-Lavoie-Roux, Montr\'eal H2V~0B3, Qu\'ebec, Canada}%
\affiliation{D\'epartement de physique, Universit\'e de Montr\'eal, 1375 Avenue Th\'er\`ese-Lavoie-Roux, Montr\'eal H2V~0B3, Qu\'ebec, Canada}%

\author{Eric~R.~Bittner}
\email{ebittner@central.uh.edu}
\affiliation{Department of Physics, University of Houston, Houston, Texas 77204, United~States}
\affiliation{Institut Courtois, Universit\'e de Montr\'eal, 1375 Avenue Th\'er\`ese-Lavoie-Roux, Montr\'eal H2V~0B3, Qu\'ebec, Canada}%

\author{Carlos~Silva-Acu\~{n}a}
\email{carlos.silva@umontreal.ca}
\affiliation{Institut Courtois, Universit\'e de Montr\'eal, 1375 Avenue Th\'er\`ese-Lavoie-Roux, Montr\'eal H2V~0B3, Qu\'ebec, Canada}%
\affiliation{D\'epartement de physique, Universit\'e de Montr\'eal, 1375 Avenue Th\'er\`ese-Lavoie-Roux, Montr\'eal H2V~0B3, Qu\'ebec, Canada}%

\date{\today}

\begin{abstract}
%
The homogeneous spectral linewidth associated with light-matter interactions is a fundamental descriptor of the optical properties of materials, governed by the quantum dynamics of the condensed-matter system. We discuss here that the homogeneous linewidth measured by means of two-dimensional electronic spectroscopy depends not only on the intrinsic microscopic dynamics of the material, but also on the observable through which those dynamics are projected onto the measurement. In this Perspective, we develop a unified framework showing that identical microscopic dynamics can yield different experimentally inferred dephasing times because different detection operators project different sectors of the nonequilibrium dynamics. 
For coherent emitted-field measurements, the observed linewidth largely retains its conventional connection to the optical coherence time $T_2$. By contrast, in population-detected modalities such as photoluminescence, photocurrent, and other action-detected two-dimensional spectroscopies, the apparent linewidth can additionally encode excited-state population redistribution dynamics, leading naturally to an effective coherence time $T_{2,\mathrm{eff}}$. Using a coupled-mode model propagated under a common Liouvillian, we show that identical microscopic dynamics yield distinct apparent dephasing times when projected onto coherent-emission and population-derived observables. The detection observable is therefore not merely part of the experimental implementation, but determines what dynamical information remains experimentally observable and how homogeneous linewidths should be interpreted as materials descriptors.
\end{abstract}

\maketitle

\section{Detection defines dephasing\label{sec:I}}
The optical response of molecular and condensed-matter systems is governed by the timescale over which electronically excited states retain the phase memory imposed by the driving electromagnetic wavepacket. The decay of this phase coherence defines the homogeneous spectral linewidth---the intrinsic width of an optical transition, inversely proportional to the coherence time and revealed by ultrafast dephasing dynamics. Homogeneous lineshapes therefore encode the fundamental quantum dynamics underlying the optical transition and provide a direct connection between microscopic fluctuations, scattering processes, and macroscopic optical function. Dephasing is often treated as an intrinsic material property determined by the system Hamiltonian and its coupling to the surrounding environment. Here we argue that, while the underlying microscopic dynamics are indeed intrinsic to the material, the experimentally inferred linewidth reflects the dynamics of the projected observable rather than the microscopic dynamics alone. In other words, the experiment does not reveal the complete density-matrix dynamics; it probes its projection onto a detection observable.

\begin{figure*}[tbh]
\centering
\includegraphics[width=0.95\textwidth]{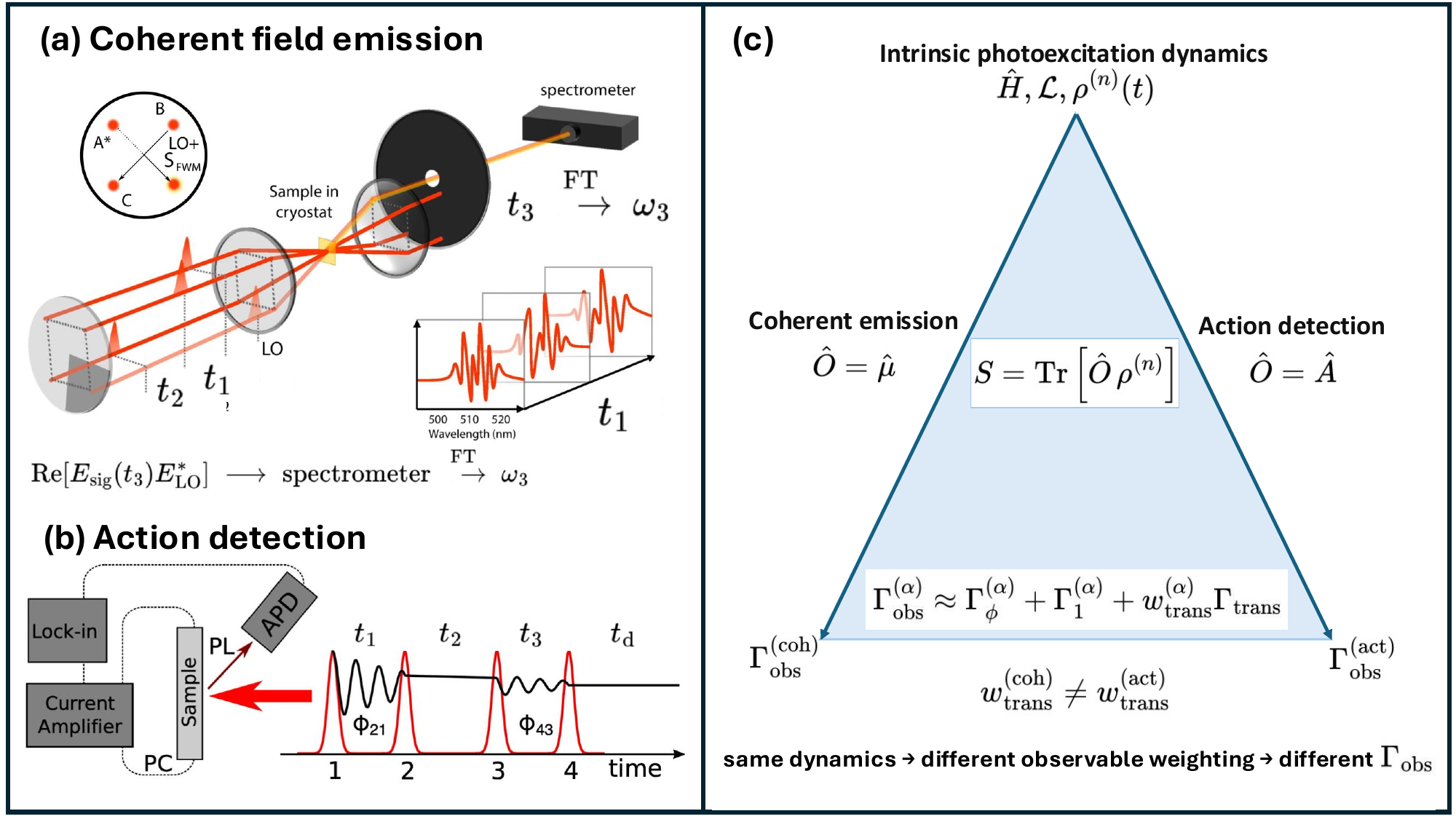}
\caption{
\textbf{Detection defines what linewidth means in multidimensional spectroscopy}.
\textbf{(a)} Schematic of a phase-matched coherent emitted-field 2DES experiment,
in which three pulses generate a third-order nonlinear polarization that
radiates a signal field detected by heterodyne interference with a local
oscillator (LO). The signal is parameterized by three time intervals:
the first coherence time \(t_1\), the population waiting time \(t_2\), and the
detection-time coherence \(t_3\), corresponding to the evolution of a
radiating optical coherence. The emitted signal field is detected by heterodyne interference
with a local oscillator, yielding a time-domain signal proportional
to the field amplitude, which is Fourier transformed (FT) with respect
to \(t_3\) to obtain the response along emission frequency \(\omega_3\).
\textbf{(b)} Schematic of a phase-tagged, population-detected (action) 2DES
experiment. Four pulses generate a nonlinear population, which is
subsequently projected onto an action observable such as
photoluminescence (PL) or photocurrent (PC) and, if applicable, integrated over a detection gate time \(t_d\). The signal is likewise parameterized
by three intervals \(t_1\), \(t_2\), and \(t_3\), where \(t_1\) and
\(t_2\) retain their usual meanings, while the final interval \(t_3\)
corresponds to the delay preceding population preparation by the final
interaction, after which the signal is projected onto
the action observable \(\hat{O}=\hat{A}\). Phase modulation of the pulse pairs allows isolation of
rephasing and nonrephasing contributions by lock-in detection.
\textbf{(c)} Conceptual hierarchy illustrating that identical photoexcitation
dynamics, governed by a common Hamiltonian \(\hat{H}\) and dissipative environment \(\mathcal{L}\),
are projected onto different observables \(\hat{O}\). Coherent emitted-field
detection probes radiating optical coherences, while action detection
probes population-derived observables. Consequently, any experimentally inferred linewidth reflects the evolution of the projected observable rather than a unique microscopic coherence timescale.
}
\label{fig:detection_defines_dephasing}
\end{figure*}

Two-dimensional electronic spectroscopy (2DES) has transformed the study of condensed-phase molecular and materials systems by separating homogeneous and inhomogeneous broadening, resolving spectral correlations in energy and time, and providing direct access to ultrafast coherence and population dynamics~\cite{jonas2003two,cho2008coherent,ginsberg2009two,nuernberger2015multidimensional,fuller2015experimental,nardin2015multi,oliver2018recent,weng2018detection,collini20212d,biswas2022coherent,fresch2023two}. In its conventional implementation (Fig.~\ref{fig:detection_defines_dephasing}a), the measured signal arises from coherent emission of the time-varying third-order nonlinear polarization, and the resulting linewidths are naturally interpreted in terms of the optical decoherence time \(T_2\), with distinct contributions from pure dephasing and population relaxation. Over the past two decades, 2DES has expanded to include implementations that detect not the emitted field itself, but rather an incoherent materials response generated by a four-pulse sequence (Fig.~\ref{fig:detection_defines_dephasing}b)~\cite{fuller2015experimental,nardin2015multi,Tekavec2007fluorescence}. Photoluminescence (PL)-detected~\cite{Tekavec2007fluorescence,de2014two,gregoire2017excitonic,gregoire2017incoherent,draeger2017rapid,goetz2018coherent,bruder2019coherent,maly2018signatures,Maly2020Coherently,Palecek2019Potential,kalaee2019differentiation,mueller2020molecular,agathangelou2021phase,gutierrez2021frenkel,liu2022perspective,bruschi2022simulating,Jayachandran2022Fluorescence-Detected,bruschi2023unifying,bruschi2024theoretical,bolzonello2023nonlinear,Javed2024Photosynthetic,jayachandran2024cogwheel,faitz2024spectrometer,bruschi2025influence,charvatova2025spectro,rueda2026theory,li2016probing}, photocurrent-detected~\cite{li2016probing,nardin2013multidimensional,karki2014coherent,vella2016ultrafast,bakulin2016ultrafast,bian2020vibronic,bolzonello2021photocurrent,chen2021time,bargigia2022identifying,amarotti2026photocurrent}, photoinduced-absorption-detected~\cite{li2016probing}, and other action-based variants have greatly broadened the range of accessible systems, extending 2DES into increasingly complex condensed-phase architectures, including \textit{in-operando} device environments~\cite{nardin2013multidimensional,karki2014coherent,vella2016ultrafast,bakulin2016ultrafast,li2016probing,bian2020vibronic,bolzonello2021photocurrent,chen2021time,bargigia2022identifying,amarotti2026photocurrent}. In these approaches, the measured signal is not the radiated nonlinear field itself, but rather a population-derived action observable.

This distinction raises a central conceptual question: \emph{what does dephasing mean when the experiment does not directly observe radiating coherence?} In conventional phase-matched coherent spectroscopy, assuming excited-state population redistribution is negligible on the timescale of the ultrafast experiment, the homogeneous spectral linewidth \(\Gamma_2\) is directly linked to the optical phase memory time \(T_2\)~\cite{Martin2018Encapsulation,Guo2020Lineshape,Bangert2021High-resolution,biswas2022coherent,De2025Quantitative} through
\begin{equation}
\Gamma_2=\frac{\hbar}{T_2}=\Gamma_{\phi}+\Gamma_{1},
\label{eq:intro_T2}
\end{equation}
where \(\Gamma_{\phi}=\hbar/T_2^{*}\) is the pure-dephasing contribution with time constant \(T_2^{*}\), and \(\Gamma_{1}=\hbar/(2T_1)\) the contribution from population relaxation with lifetime \(T_1\). In population-detected 2DES, however, the same spectral feature may additionally reflect excited-state redistribution into the measured action channel, such as integrated PL intensity.

The central thesis of this Perspective is that the operational definition of dephasing depends on the detection operator \(\hat O_\alpha\), and therefore on the experimental detection modality \(\alpha\). This distinction follows directly from the general structure of quantum measurement in nonlinear spectroscopy, where the experimentally recorded signal is obtained as the expectation value of a chosen detection operator acting on the optically prepared density matrix,
\begin{equation}
S_{\alpha}^{(n)}(t_1,t_2,\ldots,t_n)=
\mathrm{Tr}\!\left[
\hat O_{\alpha}\,\rho^{(n)}(t_1,t_2,\ldots,t_n)
\right],
\label{eq:intro_general_signal}
\end{equation}
where \(n\) denotes the number of light-matter interactions. Different detection modalities correspond to different choices of the operator \(\hat O_\alpha\), each defining a distinct probe of the same nonequilibrium density matrix \(\rho^{(n)}\). We define the first interaction to occur at \(t_0=0\), and \(t_n\) to denote the interval between the final interaction and the measurement event. For the third-order response, \(t_1\) denotes the first coherence interval and \(t_2\) the waiting time in both experiments, while the meaning of the final interval depends on the detection scheme: for coherent emitted-field detection, \(\hat O_{\alpha}=\hat\mu\), and \(t_3\) corresponds to the evolution of a radiating optical coherence, whereas for action-detected measurements, \(\hat O_{\alpha}=\hat A\), and \(t_3\) denotes the final pulse-delay interval preceding projection onto the measured population observable (Fig.~\ref{fig:detection_defines_dephasing}).

The central question addressed here is therefore not whether different detection observables produce different nonlinear spectra---a point that is well established---but rather what dynamical information remains observable under a given measurement projection. We refer to this principle as \emph{observability} in coherent spectroscopy. Because different detection operators interrogate different sectors of the nonequilibrium dynamics, they assign different operational meanings to linewidths, cross peaks, and even the dephasing time itself. A central practical question therefore arises: \emph{Do action-detected spectroscopies measure the same homogeneous linewidth as coherent emitted-field 2DES?} The prevailing view in the ultrafast spectroscopy community has been that, provided nonlinear recombination dynamics do not substantially distort the 2D coherent lineshape, the answer is tentatively yes~\cite{gregoire2017incoherent}. Subsequent work has shown that additional population dynamics can alter lineshapes, and therefore homogeneous linewidths inferred from action detection must be modeled explicitly to enable quantitative comparison with coherent-emission 2DES~\cite{Maly2020Coherently,Jayachandran2022Fluorescence-Detected,bruschi2022simulating,bolzonello2023nonlinear,Javed2024Photosynthetic}. More fundamentally, once the measured signal depends on a population-derived observable rather than directly on the radiated field, the homogeneous linewidth no longer admits a unique observable-independent interpretation.

Here, we develop a unified framework for understanding how coherent emission and incoherent population observables encode dephasing in multidimensional spectroscopy of materials. We therefore define an observable-dependent effective coherence time
\begin{equation}
T_{2,\mathrm{eff}}^{(\alpha)}
\equiv
\frac{\hbar}{\Gamma_{\mathrm{obs}}^{(\alpha)}}.
\label{eq:intro_T2eff}
\end{equation}
Equation~\eqref{eq:intro_T2eff} should therefore be viewed as an operational definition of the coherence time associated with a specified measurement protocol rather than as a universal material constant (Fig.~\ref{fig:detection_defines_dephasing}). This framework provides a route to disentangle genuine coherence loss from population redistribution and action-channel filtering in complex molecular and condensed-matter systems. The central objective is not to compare detection modalities \textit{per se}, but rather to establish how the choice of observable defines the physical meaning of dephasing and homogeneous linewidth in 2DES.

\section{Detection observables\label{sec:II}}
The key distinction between conventional phase-matched 2DES that heterodyne-detects coherent field emission (Fig.~\ref{fig:detection_defines_dephasing}a) and phase-tagged population-based detection variants (Fig.~\ref{fig:detection_defines_dephasing}b) lies in the measurement operator acting on the nonequilibrium state generated by the pulse sequence.

In coherent detection, the measured third-order nonlinear signal following three phase-matched, time-ordered ultrafast light-matter interactions is proportional to the heterodyne interference between the emitted nonlinear field due to the time-varying third-order polarization and a local oscillator (LO), derived from an attenuated replica of the femtosecond pulsetrain,
\begin{equation}
S_{\mathrm{coh}}(t_1,t_2,t_3)
\propto
\mathrm{Im}\left[
\mathbf{E}_{\mathrm{LO}}^{*}(t)
\cdot
\mathbf{P}^{(3)}(t_1,t_2,t_3)
\right],
\label{eq:Scoh_time}
\end{equation}
where \(\mathbf{E}_{\mathrm{LO}}(t)\) is the LO field that is arranged to co-propagate with the phase-matched coherent emission, and
\begin{equation}
\mathbf{P}^{(3)}(t_1,t_2,t_3)
=
\mathrm{Tr}\!\left[
\hat{\boldsymbol{\mu}}
\rho^{(3)}(t_1,t_2,t_3)
\right]
\label{eq:P3}
\end{equation}
is the time-varying, third-order mesoscopic polarization, which is the source of coherent emission.~\footnote{We retain explicit vector notation for electromagnetic fields and macroscopic polarization where required, but suppress vector indices in operator expressions after projection onto the detection axis for notational clarity.}

In contrast, action-detected measurements project the nonequilibrium density matrix produced by a sequence of four phase-tagged femtosecond pulses, which produce a final population, onto a material action observable,
\begin{equation}
S_{\mathrm{act}}^{(4)}(t_{\mathrm d};t_1,t_2,t_3)
=
\mathrm{Tr}
\left[
\hat{A}\,
\rho^{(4)}(t_1,t_2,t_3,t_{\mathrm d})
\right],
\label{eq:Sinc}
\end{equation}
where \(\hat{A}\) is the detection operator and \(t_{\mathrm d}\) denotes the detection time gate.

The operator \(\hat{A}\) may correspond to a time-integrated action variable such as PL intensity, photocurrent, photoinduced absorption, or any other experimentally accessible population-space observable as outlined in Sec.~\ref{sec:I}. Using phase-sensitive detection with a lock-in amplifier, we measure the action-detected signal integrated over the population lifetime, which is long compared to the evolution periods \(\{t_i\}\) within the Liouville path:
\begin{align}
S_{\mathrm{act}}^{(4)}(t_3,t_2,t_1)
=
\int_0^{\infty}
S_{\mathrm{act}}^{(4)}(t_{\mathrm d};t_3,t_2,t_1)\,
d t_{\mathrm d}.
\label{eq:Sact_integrated}
\end{align}
Equation~\eqref{eq:Sact_integrated} emphasizes that action detection projects the optically prepared state through its subsequent population evolution before the signal is recorded. The integration over \(t_d\) therefore contributes to the effective observable through which population redistribution is sampled~\cite{Gellen2017Probing,Mueller2018Fluorescence-Detected,Maly2020Coherently,bolzonello2021photocurrent,bruschi2022simulating,Jayachandran2022Fluorescence-Detected}; it is not introduced here as an independent additive dephasing channel.

More generally, the measured nonlinear signal may be written as Eq.~\eqref{eq:intro_general_signal}, where \(\hat O_\alpha\) specifies the observable. Because these operators interrogate different sectors of the nonequilibrium dynamics, changing \(\hat O_\alpha\) changes the operational definition of dephasing. This rationale supports the view that dephasing is not only a dynamical property of the material Hamiltonian; it is also a property of how the system-bath quantum dynamics project onto the observable intrinsic to the experiment.

It is worth noting that the experimental scheme and strategy shown in Fig.~\ref{fig:detection_defines_dephasing}b can be implemented to realize the two measured observables discussed in this section. As depicted in this figure, it represents the projection of Eq.~\eqref{eq:Sinc} if the four-pulse sequence generates the final population on which the action observation is performed, but the fourth pulse can also be used as a LO to extract the time-varying polarization to project as in Eq.~\eqref{eq:Scoh_time}, as exemplified in Ref.~\citenum{Martin2018Encapsulation}. The collinear geometry with a phase-tagging, phase-sensitive detection strategy can be implemented for both projection schemes discussed in this Perspective, offering a powerful option for microscopy and other applications in which the phase-matching beam geometry, as in Fig.~\ref{fig:detection_defines_dephasing}a, is inconvenient.

\subsection{Observable-Dependent Dephasing Dynamics}
\label{sec:minimal_observable_projection}

The observable dependence of dephasing can be simply illustrated using a two-level system with Hamiltonian
\[
\hat H_0
=
\frac{\hbar\omega_0}{2}\sigma_z.
\]
The system undergoes identical microscopic dynamics but is subjected to different measurements, namely coherent field-emission and action-detected measurements. In the context of multidimensional spectroscopy, the expectation value of an observable operator \(\hat O\) to \(n\)-th order at time \(t\) can be expressed in the interaction representation as
\begin{equation}
\langle \hat O^{(n)}(t) \rangle
\equiv
\mathrm{Tr}\!\left[
\hat O \rho^{(n)}(t)
\right]
=
\mathrm{Tr}\!\left[
\hat O_{\mathrm I}(t)
\rho^{(n)}_{\mathrm I}(t)
\right].
\label{eq:adjoint_projection_identity}
\end{equation}
The interaction representation is equivalent to the Schr\"odinger picture but is convenient for analyzing the final measurement \(\hat O\). Because \(\rho_{\mathrm I}\) is time-independent in the absence of perturbations, we can focus on the dynamics of \(\hat O_{\mathrm I}(t)\).

Taking the Lindblad master equation
\begin{equation}
\frac{d\hat O_{\mathrm I}}{dt}
=
\frac{i}{\hbar}
[\hat H_0,\hat O_{\mathrm I}]
+
\sum_j
\left(
\hat L_j^\dagger
\hat O_{\mathrm I}
\hat L_j
-
\frac{1}{2}
\left\{
\hat L_j^\dagger
\hat L_j,
\hat O_{\mathrm I}
\right\}
\right),
\label{eq:two_level_master_equation}
\end{equation}
we consider two Lindblad operators,
\[
\hat L_1
=
\sqrt{\gamma_1}\,\sigma_-,
\qquad
\hat L_2
=
\sqrt{\gamma_{\phi}/2}\,\sigma_z,
\]
that characterize dissipation due to population relaxation and energy-gap fluctuations, respectively. The population-relaxation rate is \(\gamma_1=1/T_1\), and \(\gamma_\phi=1/T_2^{*}\) is the pure-dephasing rate. The raising and lowering operators are \(\sigma_{\pm}=\sigma_x\pm i\sigma_y\), and \(\sigma_{x,y,z}\) are the Pauli matrices.

For coherent field-emission detection, the measurement operator is the dipole operator \(\hat\mu\propto\sigma_x\), which satisfies
\begin{equation}
\frac{d\hat\mu}{dt}
=
i\omega_0
\left(
\sigma_+ - \sigma_-
\right)
-
\left(
\frac{\gamma_1}{2}
+
\gamma_{\phi}
\right)
\hat\mu.
\label{eq:dipole_evolution}
\end{equation}
The dephasing of the polarization is characterized by the second term on the right-hand side of Eq.~\eqref{eq:dipole_evolution}, with decay rate
\begin{equation}
\gamma_{\mathrm{coh}}
=
\frac{\gamma_1}{2}
+
\gamma_{\phi},
\label{eq:dipole_op_decay_rate}
\end{equation}
consistent with the linewidth specified in Eq.~\eqref{eq:intro_T2}.

For action detection, the relevant operator is the excited-state projection operator
\[
\hat A
=
\frac{\sigma_z+I}{2},
\]
with evolution
\begin{equation}
\frac{d\hat A}{dt}
=
-\gamma_1\hat A.
\label{eq:action_operator_evolution}
\end{equation}
The relaxation operator \(\hat L_1\) affects the excited-state population \(\langle\hat A\rangle\), whereas pure dephasing associated with energy-gap fluctuations makes no contribution. The population decay rate is therefore
\begin{equation}
\gamma_{\mathrm{pop}}
=
\gamma_1.
\label{eq:population_op_decay_rate}
\end{equation}

Equations~\eqref{eq:dipole_op_decay_rate} and \eqref{eq:population_op_decay_rate} demonstrate the central observable dependence in its minimal form. The dynamics contain the same population relaxation and the same pure dephasing, but the dipole operator and population projector sample different Liouville-space sectors, leading to different decay rates. The coherent polarization is sensitive to both processes, whereas the population observable is blind to pure dephasing. Thus, the operator-specific decay constants differ even though the Hamiltonian and dissipative dynamics are identical.

Furthermore, the minimal model provides a qualitative estimate of the homogeneous linewidths associated with the two spectroscopic approaches. When pure dephasing prevails, \(\gamma_{\phi}>\gamma_1/2\), the homogeneous linewidth of the coherent field-emission spectrum is broader than that of the action-detected spectrum. Conversely, when population relaxation dominates, the action-detected spectrum exhibits a broader homogeneous lineshape.

This example establishes the organizing principle used throughout the Perspective: experimentally inferred dynamical quantities are determined jointly by the underlying quantum dynamics and by the observable through which those dynamics are measured. In practice, multidimensional spectroscopy deals with systems containing more states that involve population redistribution, competing decay channels, and state-dependent action yields, as discussed in Sec.~III. The analysis of the two-level model should therefore be viewed as a minimal operator-level illustration rather than as a complete description of dephasing in 2D spectroscopies.

\subsection{Shared pathway selectivity across detection modalities}

We return to the question: \emph{How are the two experimental modalities in Fig.~\ref{fig:detection_defines_dephasing} fundamentally different?} An important point of commonality between coherent emitted-field and population-detected 2DES is that both detection schemes retain access to selected Liouville pathway classes~\cite{biswas2022coherent}. In phase-matched three-pulse experiments, wavevector selection geometrically isolates rephasing and nonrephasing contributions through their distinct phase-matching conditions. In collinear phase-tagged population detection, the same pathway classes can be isolated through appropriate phase cycling or phase demodulation, enabling direct construction of rephasing-like and nonrephasing-like responses from the same underlying nonequilibrium quantum dynamics. For either modality, one may formally write
\begin{equation}
S_{\alpha}^{(R/NR)}
=
\mathrm{Tr}
\left[
\hat O_\alpha
\rho_{R/NR}^{(n)}
\right],
\label{eq:RNR_projection}
\end{equation}
where \(\rho_{R/NR}^{(n)}\) denotes the selected rephasing or nonrephasing density-matrix component.

This shared pathway selectivity is conceptually important because, once a pathway class has been isolated, the detected signal still depends on the observable operator \(\hat O_\alpha\) through which the trajectory is read out. Rephasing and nonrephasing spectra may therefore remain formally analogous across modalities while still encoding different operational meanings for linewidths, cross peaks, and effective coherence times.

\subsection{Pathway-summed pump-probe implementations}

A useful point of comparison is provided by pump-probe implementations of 2DES based on passively phase-locked pump pulse pairs generated, for example, with birefringent wedges and detected through white-light probe self-heterodyning~\cite{rehault2014two}. In this geometry, the broadband white-light probe pulsetrain serves as an intrinsic local oscillator for the third-order differential transmission signal, preserving high phase stability and experimental simplicity while directly encoding spectral correlations along the pump coherence and probe detection axes.

Unlike the explicitly pathway-resolved coherent phase-matched and phase-tagged population-detected modalities discussed above, however, this implementation typically reports a pathway-summed correlation function rather than separated Liouville components~\cite{rehault2014two}. Unless additional phase cycling, pulse-order discrimination, or interferometric separation is introduced~\cite{Zhu2017Broadband,Farrell2022Phase,Luttig2023High-order,Cai2024Extracting,Timmer2025Disentangling}, the measured spectrum is more naturally viewed as the total response,
\[
S_{\mathrm{tot}}
=
S_R+S_{NR}.
\]

This distinction is important for the present Perspective because the linewidth and lineshape nuances discussed here are most rigorously defined when specific Liouville pathway classes are explicitly resolved. The observable dependence emphasized throughout this article therefore applies to the total projected response in an averaged sense, rather than to individually resolved sectors~\cite{Tokmakoff:2000aa}. This does not diminish the utility of pump-probe 2D methods, but it does place the present discussion of effective observed linewidths in the narrower context of explicitly pathway-resolved 2DES measurements; it is our strong statement here that pathway-specific lineshape analysis is one of the most compelling advantages of 2DES~\cite{Tokmakoff:2000aa}, and pathway-summed 2DES is a useful way to enhance general pump-probe methods, but does not realize the full potential of 2DES. The hierarchy of pathway resolution provides the natural starting point for asking how linewidths and apparent dephasing times inherit distinct observable-dependent contributions.

\subsection{Extension to Higher-Order Response: Accessing Many-Body Correlations}

While the discussion above has focused on third-order response, the role of the detection operator in selecting experimentally accessible dynamics extends naturally to higher-order nonlinear processes. Throughout this Perspective we distinguish the quantum order of the detected coherence, one-quantum or two-quantum, from the perturbative order of the nonlinear optical response. Although the leading contribution to the two-quantum signal considered here is fifth order, higher-order nonlinear responses may also contribute, depending on excitation conditions and detection geometry. Two-quantum coherences may also be accessed at third order in appropriate phase-matching directions; consequently, the designation ``two-quantum'' specifies the coherence sector being interrogated rather than the perturbative order of the measured response.

At the fifth-order level, the measured signal remains of the general form
\begin{equation}
S_{\alpha}^{(5)}(t_1,\ldots,t_5)
\propto
\mathrm{Tr}\!\left[
\hat O_{\alpha}\,
\rho^{(5)}(t_1,\ldots,t_5)
\right],
\end{equation}
but now reflects the evolution of higher-rank multi-time correlations in the system. As in the third-order case, the key distinction lies in the detection operator \(\hat O_{\alpha}\).

Experimentally, the two-quantum signal dominated by its leading fifth-order response in action detection can be isolated via higher-harmonic demodulation, which imposes selection rules in Liouville space. In our work on Frenkel biexcitons~\cite{gutierrez2021frenkel}, demodulation at harmonics corresponding to two interactions with an initial pulse pair selects pathways that generate two-quantum coherences, which are subsequently projected onto populations and converted back to single-quantum coherences by later interactions. This sequence defines a specific population--coherence--population conversion pathway that is directly encoded in the measured observable.

At this level, the signal encodes not only the presence of multi-exciton states, but also the dynamics that interconvert coherences and populations within these manifolds. In analogy to the decomposition of \(\Gamma_{\mathrm{obs}}\) at third order, one can interpret the fifth-order multi-quantum signal in terms of distinct physical processes, including correlated dephasing, population transfer within the two-exciton manifold, and relaxation to single-exciton states. These processes contribute with different weights depending on \(\hat O_{\alpha}\).

Complementary information is obtained in coherent field-emission 2DES, where phase-matched detection isolates the leading nonlinear polarization pathways contributing to the selected phase-matching direction. In the present implementation, the dominant contribution to the two-quantum coherent signal is fifth order, although higher-order nonlinear responses may also contribute under sufficiently strong excitation conditions. In semiconductor quantum wells, this approach has been used to directly resolve a two-quantum signal dominated by its leading fifth-order response (\textit{e.g.}, Turner and Nelson~\cite{turner2010coherent}), providing access to coherent multi-exciton correlations that are not projected onto populations.

Formally, the underlying response involves higher-order dipole correlation functions,
\[
\langle
\hat{\mu}(t_5)
\hat{\mu}(t_4)
\hat{\mu}(t_3)
\hat{\mu}(t_2)
\hat{\mu}(t_1)
\rangle,
\]
but the experimentally accessible information is determined by how these correlations are projected by the detection operator. As a result, coherent emission and action detection select different components of the same many-body dynamics.

Thus, extending to the dominant \(\chi^{(5)}\) contribution preserves the central conclusion established at third order: the observed signal does not directly reflect the intrinsic dynamics alone, but rather the dynamics as filtered by the detection observable. Higher-order spectroscopy therefore provides a more intricate route to selectively accessing many-body correlations, with detection modality acting as a control parameter for which processes are emphasized or suppressed. This opens new opportunities to significantly advance our understanding of multi-quasiparticle physics~\cite{koch2026biexcitons}.

\section{Detection defines 2DES lineshapes\label{sec:III}}
The remainder of this Perspective develops the basic idea in three steps. We first derive an observable-dependent decomposition of the measured linewidth, then illustrate the concept using a minimal Liouville-space model with fixed microscopic dynamics, and finally compare the predictions with representative experimental observations in conjugated polymers.

\subsection{The observed homogeneous linewidth}

Building on the observable-specific definition of
\(T_{2,\mathrm{eff}}^{(\alpha)}\) introduced in
Eq.~\eqref{eq:intro_T2eff}, we decompose the measured linewidth into
contributions arising from coherence loss, population relaxation and
excited-state redistribution. The
normalized time-domain signal envelope is written as
\begin{equation}
C_{\mathrm{obs}}(t)
\equiv
\frac{S(t)}{S(0)}
\approx
C_{\phi}(t)\,
C_{1}(t)\,
C_{\mathrm{trans}}^{(\alpha)}(t),
\label{eq:Cobs_factorized}
\end{equation}
where \(C_{\phi}(t)\) describes pure phase randomization,
\(C_{1}(t)\) population relaxation,
and \(C_{\mathrm{trans}}(t)\) excited-state population redistribution. The factorization is not intended as an exact microscopic decomposition of the Liouvillian dynamics. Rather, it provides an effective phenomenological framework for organizing the principal physical processes that contribute to the experimentally observed linewidth.

If these processes are approximately separable and exponential over
the experimental window,
\begin{equation}
C_{\mathrm{obs}}(t)
\approx
\exp\!\left[
-\frac{
(\Gamma_\phi+\Gamma_1+\Gamma_{\mathrm{trans}})
t}{\hbar}
\right].
\label{eq:Cobs_total}
\end{equation}
This decomposition is introduced as the minimal phenomenological framework required to distinguish contributions arising from coherence decay, population relaxation, population redistribution, and the detection channel. It is not intended as a microscopic derivation applicable to arbitrary open quantum systems. The effective linewidth is then
\begin{equation}
\Gamma_{\mathrm{obs}}
\approx
\Gamma_{\phi}
+
\Gamma_{1}
+
\Gamma_{\mathrm{trans}}.
\label{eq:Gammaobs}
\end{equation}
Here again, the factorization is not intended as an exact microscopic decomposition of the Liouvillian. Rather, it provides a phenomenological framework for organizing the principal physical processes contributing to the experimentally observed linewidth. In general, non-Markovian fluctuations, correlated vibronic dynamics, and nonlinear population mixing may lead to nonexponential dynamics that do not admit a unique separation. These regimes motivate the discussion in Sec.~\ref{sec:V}, where incoherent population mixing and pseudo-dephasing are examined as experimentally important examples of the limitations of this phenomenological picture.

A detection-dependent generalization is
\begin{equation}
\Gamma_{\mathrm{obs}}^{(\alpha)}
\approx
\Gamma_{\phi}^{(\alpha)}
+
\Gamma_{1}^{(\alpha)}
+
w_{\mathrm{trans}}^{(\alpha)}\Gamma_{\mathrm{trans}},
\label{eq:Gamma_general_weighted}
\end{equation}
where \(\alpha\) denotes the detection modality. The first two terms describe intrinsic optical decoherence. The latter term describe how the chosen observable projects the common nonequilibrium dynamics onto the experimentally recorded signal: \(w_{\rm trans}^{(\alpha)}\Gamma_{\rm trans}\) accounts for observable-weighted population redistribution. The weighting coefficients encode how a given experimental observable samples the underlying nonequilibrium dynamics rather than representing intrinsic properties of the material. Coherent emission samples transfer through its effect on evolving optical coherences, whereas action detection samples populations after redistribution and observable filtering, leading to distinct weights \(w_{\mathrm{trans}}^{(\alpha)}\).

The observable dependence enters through \(w_{\mathrm{trans}}^{(\alpha)}\), which describes how population redistribution under the common Liouvillian is projected onto the experimentally recorded signal. For action detection, this weighting incorporates state-dependent action yields and any post-excitation population evolution sampled by the detection protocol. These effects therefore modify the observable projection of redistribution rather than constituting an independent additive linewidth channel.

The inferred dephasing time \(T_{2,\mathrm{eff}}^{(\alpha)}\) is therefore an operational property of the measurement protocol acting on common microscopic dynamics, rather than a property of the system Hamiltonian alone.

This detection dependence is summarized schematically in
Fig.~\ref{fig:detection_defines_dephasing}c, where a common set of
microscopic dynamics is projected onto distinct observables, yielding
different operational definitions of the homogeneous linewidth.

Beyond linewidths, the measurement modifies other aspects of the 2DES
lineshape. Cross-peaks in coherent emission reflect phase-coherent
pathways, whereas in population detection they are additionally shaped
by redistribution and observable weighting. As a result, amplitudes and
temporal evolution can differ even when the underlying dynamics are
identical. To make this decomposition operational, we now construct a model that represents a fixed physical system probed by both experimental protocols.

\subsection{Model: Common dynamics, distinct observables}

To isolate the role of the detection observable independently of material-specific disorder, excitonic microstructure, and other system-dependent complexities, we consider a minimal model in which the Hamiltonian and
dissipative dynamics are held fixed while only the detection operator is
changed, as illustrated schematically in
Fig.~\ref{fig:detection_defines_dephasing}c. The purpose of the simulations is not to validate the phenomenological decomposition itself, but to demonstrate that identical microscopic dynamics produce distinct experimentally inferred linewidths when projected onto different observables. The objective is therefore not to reproduce the full complexity of any particular condensed-matter system, but to isolate the observable dependence of the inferred linewidth under otherwise identical microscopic dynamics. Material-specific effects such as static disorder, energetic heterogeneity, and correlated transport can subsequently be incorporated within the same observable-projection framework without altering its central conceptual premise.

The model is naturally formulated in Liouville space, where the density matrix
evolves under a common Liouvillian and different detection operators
project distinct signals from the same underlying trajectory. We consider
two weakly anharmonic coupled modes,
\begin{equation}
\hat H_0
=
\sum_{i=1}^{2}
\left[
\omega_i \hat b_i^\dagger \hat b_i
+
\frac{\Delta_i}{2}
\hat b_i^\dagger \hat b_i^\dagger \hat b_i \hat b_i
\right]
+
J
\left(
\hat b_1^\dagger \hat b_2
+
\hat b_2^\dagger \hat b_1
\right),
\label{eq:H_model}
\end{equation}
where \(\hat b_i^\dagger\) and \(\hat b_i\) create and annihilate
excitations in mode \(i\), \(\omega_i\) are the mode frequencies,
\(\Delta_i\) are weak anharmonicities, and \(J\) is the intermode
coupling. We consider a weakly anharmonic excitonic manifold that ensures a nontrivial nonlinear response while preserving a transparent normal-mode structure, for which the excited-state absorption transition is only slightly displaced from the corresponding stimulated-emission pathway. This regime minimizes artificial linewidth differences arising solely from pathway separation and therefore provides a more stringent test of whether observable-dependent detection alone modifies the experimentally inferred homogeneous linewidth.

The light-matter interaction is treated perturbatively in the
Coulomb gauge,
\begin{equation}
\hat H_{\mathrm{int}}(t)
=
-\hat{\boldsymbol{\mu}}\cdot\mathbf{E}(t),
\label{eq:Hint_model}
\end{equation}
with dipole operator
\begin{equation}
\hat \mu
=
\sum_{i=1}^{2}
\mu_i
\left(
\hat b_i+\hat b_i^\dagger
\right),
\label{eq:mu_model}
\end{equation}
where \(\hat{\mu}\equiv\hat{\boldsymbol{\mu}}\cdot\mathbf{e}_{\mathrm{det}}\), with \(\mathbf{e}_{\mathrm{det}}\) being the detection vector that encodes the measurement operator defining the observable signal.

The reduced density matrix evolves according to
\begin{equation}
\begin{split}
\frac{d\rho}{dt}
=
-&\frac{i}{\hbar}
\left[
\hat H_0+\hat H_{\mathrm{int}}(t),\rho
\right]\\
&+
\mathcal L_{\phi}[\rho]
+
\mathcal L_{\mathrm{rel}}[\rho]
+
\mathcal L_{\mathrm{trans}}[\rho]
+
\mathcal L_{\mathrm{rad}}[\rho],
\end{split}
\label{eq:master_model}
\end{equation}
where the dissipators describe pure dephasing, population relaxation,
population redistribution, and radiative decay, respectively. The simulations therefore establish the existence of observable-dependent linewidths under identical microscopic dynamics, independently of the quantitative details of any particular material system.

Numerical simulations are performed using our quantum dynamics package \texttt{QuDPy}~\cite{shah2023qudpy}, modeling the system described by Eq.~\eqref{eq:H_model}, truncated to three levels per mode. This nine-dimensional Hilbert space provides a minimal description of the ground \(|g\rangle\), one-quantum \(\{|a\rangle\}\), and two-quantum \(\{|f\rangle\}\) manifolds required for both coherent response and population redistribution dynamics. Both the phase-matched emitted-field response and the phase-tagged PL-detected signal are propagated under the same common Liouvillian as in Fig.~\ref{fig:detection_defines_dephasing}c, and the corresponding signals are determined according to Eqs.~\eqref{eq:Scoh_time} and~\eqref{eq:Sinc}.

Pure dephasing is modeled as
\begin{equation}
\mathcal L_{\phi}[\rho]
=
\sum_a
\gamma_{\phi,a}
\left(
\hat P_a \rho \hat P_a
-
\frac{1}{2}
\left\{
\hat P_a,\rho
\right\}
\right),
\label{eq:Lphi_model}
\end{equation}
where \(\hat P_a=|a\rangle\langle a|\) projects onto excited state
\(a\), and \(\gamma_{\phi,a}\) is the corresponding dephasing rate.

Population relaxation is described as a cascade from the two-quantum
manifold \(\{|f\rangle\}\) into the one-quantum manifold \(\{|a\rangle\}\),
followed by decay to the ground state \(|g\rangle\),
\begin{equation}
\begin{split}
\mathcal L_{\mathrm{rel}}[\rho]
&=
\sum_{f,a}
k^{(2\to1)}_{fa}
\left(
\hat L_{af}\rho \hat L_{af}^{\dagger}
-\frac12\{\hat L_{af}^{\dagger}\hat L_{af},\rho\}
\right)\\
&+
\sum_a
k^{(1\to0)}_a
\left(
\hat L_{ga}\rho \hat L_{ga}^{\dagger}
-\frac12\{\hat L_{ga}^{\dagger}\hat L_{ga},\rho\}
\right),
\end{split}
\label{eq:Lrel_model}
\end{equation}
with \(\hat L_{af}=|a\rangle\langle f|\) and
\(\hat L_{ga}=|g\rangle\langle a|\). Although these jump operators are
not Hermitian, the Lindblad form guarantees preservation of both
Hermiticity and trace of \(\rho\).

Population redistribution within the one-quantum manifold is described by
\begin{equation}
\begin{split}
\mathcal L_{\mathrm{trans}}[\rho]
&=
\sum_{a\neq b}
k_{b\to a}
\left(
\hat T_{ba}\rho \hat T_{ba}^{\dagger}
-\frac12\{\hat T_{ba}^{\dagger}\hat T_{ba},\rho\}
\right)\\
&+
k_{a\to b}
\left(
\hat T_{ab}\rho \hat T_{ab}^{\dagger}
-\frac12\{\hat T_{ab}^{\dagger}\hat T_{ab},\rho\}
\right),
\end{split}
\label{eq:Ltrans_model}
\end{equation}
with \(\hat T_{ba}=|b\rangle\langle a|\) and
\(\hat T_{ab}=|a\rangle\langle b|\). This term captures processes such as
energy transfer or exciton migration between excited states. For
thermally activated reversible exchange, the rates satisfy detailed
balance,
\begin{equation}
\frac{k_{b\to a}}{k_{a\to b}}
=
\exp\left[
-\frac{\hbar(\omega_a-\omega_b)}{k_{\mathrm B}T}
\right],
\end{equation}
so that downhill transfer is favored according to the Boltzmann
population ratio.

Radiative decay from the one-quantum manifold is given by
\begin{equation}
\begin{split}
\mathcal L_{\mathrm{rad}}[\rho]
&=
\sum_a
k_{\mathrm{rad},a}
\left(
\hat R_a \rho \hat R_a^{\dagger}
-\frac12\{\hat R_a^{\dagger}\hat R_a,\rho\}
\right),\\
\hat R_a &= |g\rangle\langle a|,
\end{split}
\label{eq:Lrad_model}
\end{equation}
where \(k_{\mathrm{rad},a}\) is the spontaneous emission rate of state
\(|a\rangle\). These radiative channels define the intrinsic emission
processes, while the detection operator determines how strongly each
channel contributes to the measured signal.

\begin{figure*}[ht]
\centering
\includegraphics[width=\linewidth]{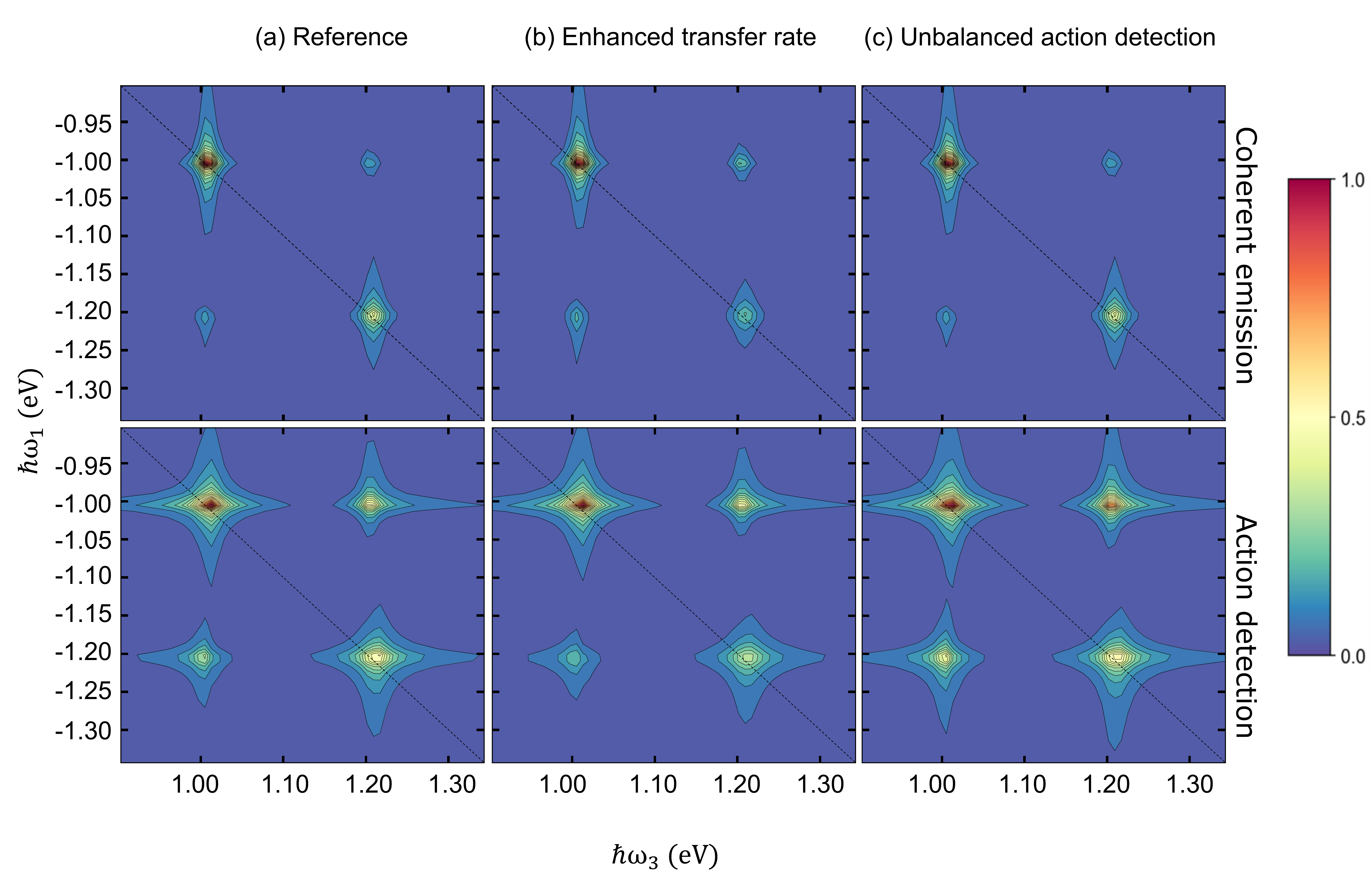}
\caption{
\textbf{Simulated 2DES amplitude rephasing spectra in a rephasing Liouville pathway, illustrating the effect of detection modality on the observed nonlinear lineshape} in a low-temperature regime of Eq.~\eqref{eq:Ltrans_model} (no uphill excited-state population transfer), and with \(\hbar\omega_a=1.0\)\,eV, \(\hbar\omega_b=1.2\)\,eV, and \(\hbar\Delta=0.01\)\,eV. Transition dipole moments were normalized such that \(|\hat{\boldsymbol{\mu}}_1|=|\hat{\boldsymbol{\mu}}_2|=1\).
The top row shows field-emission-detected spectra (\(\hat O_{\mathrm{coh}}=\hat\mu\)), while the bottom row shows action-detected spectra (\(\hat O_{\mathrm{act}}=\hat A\)) calculated from the same underlying excitonic Hamiltonian and dissipative dynamics, Eq.~\eqref{eq:master_model}.
The columns correspond to three parameter regimes:
\textbf{(a)} the leftmost column is the baseline reference case with microscopic parameters \(\hbar\gamma_{\phi}=10\)\,meV for both modes, \(\hbar k^{(1\to0)}_a=1\)\,meV and \(k^{(2\to1)}=2k^{(1\to0)}\) for both modes, \(\hbar k_{b\to a}=5\)\,meV, \(\hbar k_{\mathrm{rad}}=0.1\)\,meV for both modes, and \(\eta_1=\eta_2\) in the action operator;
\textbf{(b)} the middle column represents a case with modified population-transfer rate \(\hbar k_{b\to a}=10\)\,meV, with all other parameters as in the baseline simulation;
\textbf{(c)} the rightmost column uses the same parameters as the baseline simulation but with unbalanced action-detection weights \(\eta_1=2\eta_2\). These parameter values were chosen to avoid the strong-coupling regime; the rates and coupling strength are varied as described in Table~\ref{tab:Simulation_Linewidths}.
}
\label{fig:simulated_amplitude_spectra}
\end{figure*}

Within this common dynamical framework, we compare two observables. In
coherent emitted-field detection, the signal is obtained from the
third-order polarization \(\mathbf{P}^{(3)}(t_3,t_2,t_1)\), Eq.~\eqref{eq:P3},
which directly probes the evolution of optical coherences. In contrast,
in PL-detected 2DES the signal is obtained from a
population-weighted action operator acting on the emissive manifold,
\begin{equation}
\hat A_{\mathrm{PL}}
=
\sum_a
\eta_a |a\rangle\langle a|.
\label{eq:APL_general_model}
\end{equation}
From the standpoint of quantum measurement theory, changing from coherent detection to action detection corresponds only to replacing the final observable while leaving the Liouvillian propagation unchanged. For the present two-mode model, Eq.~\eqref{eq:APL_general_model} reduces to
\begin{equation}
\hat A_{\mathrm{PL}}
=
\eta_1 |e_1\rangle\langle e_1|
+
\eta_2 |e_2\rangle\langle e_2|.
\label{eq:APL_twolevel_model}
\end{equation}
The coefficients \(\eta_a\) encode the emissive yield and detection
weight of each state, so that the observable reflects both intrinsic
radiative decay and any population redistribution preceding emission.

The key point is that the Hamiltonian and all dissipative processes are identical in the two cases; only the observable differs. This distinction is reflected experimentally in 2DES of silicon-vacancy centers in diamond, where PL- and heterodyne-detected measurements yield markedly
different spectra~\cite{smallwood2021hidden}. In that system, PL
detection effectively post-selects a subset of radiatively efficient
emitters, whereas heterodyne detection probes a broader ensemble,
leading to different linewidths and coherence times.
Within the present framework, this corresponds to observable-dependent projection of a common dynamical ensemble. The work of Smallwood~\emph{et al.}~\cite{smallwood2021hidden} therefore illustrates an important general principle: changing the observable changes the experimentally accessible projection of the underlying dynamics without changing the underlying Hamiltonian itself, as expressed in Fig.~\ref{fig:detection_defines_dephasing}. Here we extend this principle beyond ensemble selection, showing that even for a fixed set of microscopic dynamics, different observables can weight coherence, population redistribution, and relaxation processes differently.
As a result, the measured homogeneous linewidth becomes an
observable-dependent quantity, rather than a unique property of the
underlying system.

We show analytically in the Supplemental Material that a two-level realization of the model of Eq.~\eqref{eq:H_model}, propagated according to Eq.~\eqref{eq:master_model}, results in a deviation of the observed homogeneous linewidth for resonance \(b\) upon action detection, in the low-temperature limit \(k_{\mathrm B}T\ll\hbar(\omega_b-\omega_a)\),
\begin{equation}
\Delta \Gamma_{\mathrm{obs}}^{(b)}
\simeq
k_{b\to a}
\left(
1-\frac{\eta_a}{\eta_b}
\right).
\label{eq:low-temp}
\end{equation}
We assume \(\omega_b>\omega_a\), such that in the low-temperature limit
population transfer is unidirectional and downhill, \(b\to a\). In the high-temperature limit \(k_{\mathrm B}T\gg\hbar(\omega_b-\omega_a)\), detailed balance gives \(k_{b\to a}\simeq k_{a\to b}\simeq k_{\mathrm{ex}}/2\), so that the initial-slope result becomes
\begin{equation}
\Delta \Gamma_{\mathrm{obs}}^{(b)}
=
\frac{k_{\mathrm{ex}}}{2}
\left(
1-\frac{\eta_a}{\eta_b}
\right),
\label{eq:high-temp}
\end{equation}
with \(k_{\mathrm{ex}}\equiv k_{b\to a}+k_{a\to b}\). Population transfer contributes to the initial observed linewidth through the downhill component of the thermally parametrized exchange rate, weighted by the detection ratio \(\eta_a/\eta_b\), and vanishes identically for \(\eta_a=\eta_b\). At finite evolution times, the full bidirectional exchange dynamics are retained by the Liouvillian propagation.

We now model the 2DES lineshapes numerically~\cite{shah2023qudpy} to explore this parameter landscape. The numerical simulations presented below verify that this transport-induced observable divergence persists under experimentally realistic pure-dephasing conditions and increases systematically with population transport, demonstrating that it is not an artifact of lifetime-limited dynamics.

\subsection{2DES simulated lineshapes}

\begin{figure}[tb]
\centering
\includegraphics[width=8.5cm]{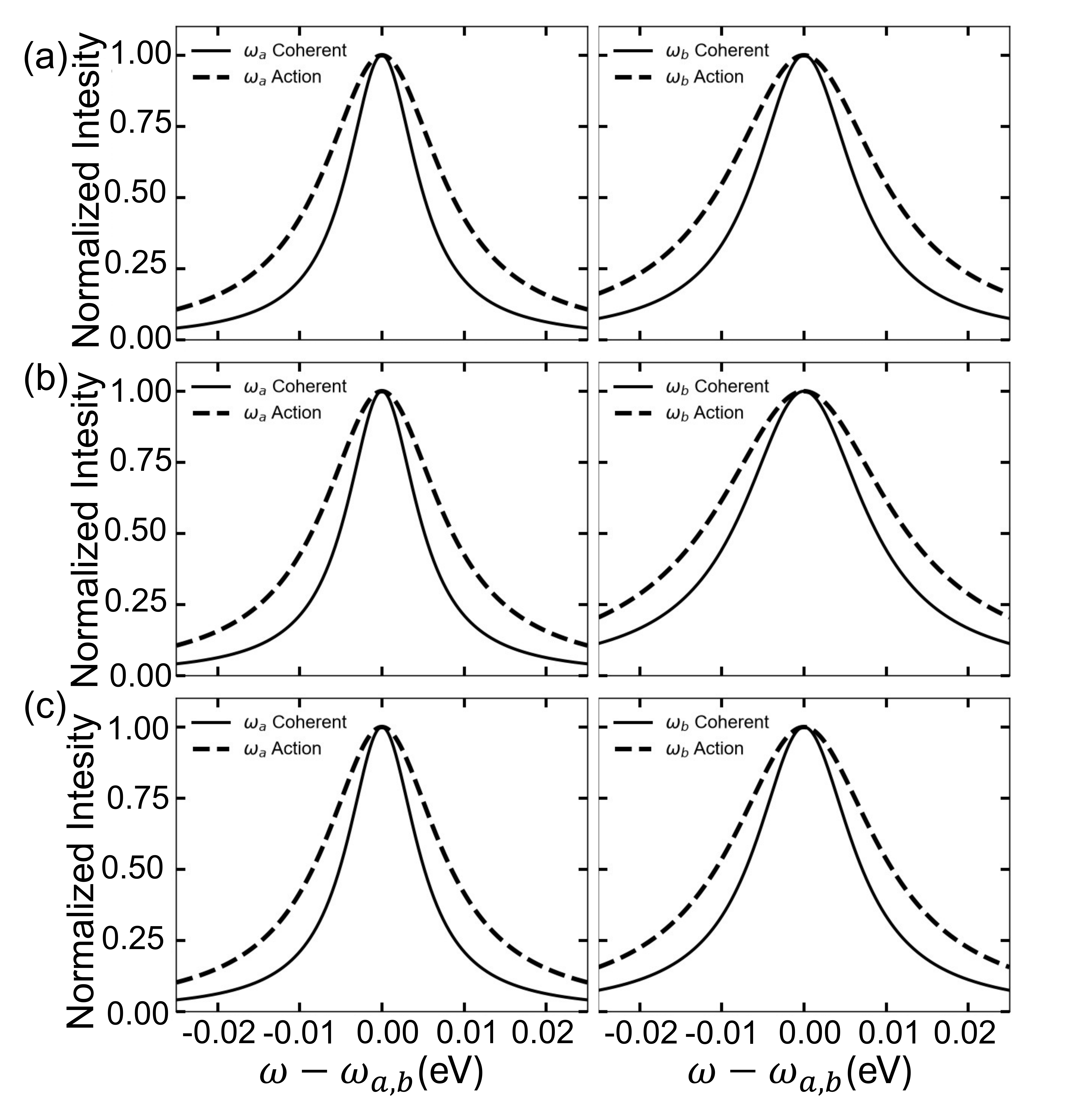}
\caption{
\textbf{Antidiagonal cuts through the simulated 2DES quantify the detection dependence of the apparent homogeneous linewidth}. Shown are cuts for
\textbf{(a)} the reference simulations [Fig.~\ref{fig:simulated_amplitude_spectra}(a)],
\textbf{(b)} the enhanced excited-state transfer simulations [Fig.~\ref{fig:simulated_amplitude_spectra}(b)], and
\textbf{(c)} the unbalanced detection weights [Fig.~\ref{fig:simulated_amplitude_spectra}(c)].
}
\label{fig:antidiagonal_cuts}
\end{figure}

Figure~\ref{fig:simulated_amplitude_spectra} compares simulated rephasing 2DES amplitude spectra generated from the same underlying quantum dynamics of Eq.~\eqref{eq:master_model}, but projected onto two different detection observables. The top row shows coherent emitted-field detection, while the bottom row shows action detection. In the reference case [Fig.~\ref{fig:simulated_amplitude_spectra}(a)], both detection schemes recover the same basic spectral structure, with diagonal features associated with the two excitonic resonances and cross peaks reflecting correlations between them. Despite the common Hamiltonian and Liouvillian dynamics, however, the two observables produce markedly different lineshapes. The action-detected resonances are systematically broader than their coherent-emission counterparts, demonstrating that the experimentally inferred homogeneous linewidth depends on the observable through which the common nonequilibrium dynamics are projected. The relative cross-peak amplitudes also differ substantially between the two detection modalities, with the action-detected spectrum exhibiting considerably stronger cross-peak intensity relative to the diagonal features. Thus, the observable dependence extends beyond linewidth broadening to the experimentally accessible manifestation of spectral correlations.

Increasing the downhill population-transfer rate \(k_{b\to a}\) [Fig.~\ref{fig:simulated_amplitude_spectra}(b)] redistributes spectral weight within both detection schemes, modifying the relative intensities of the diagonal and cross-peak features. In particular, the higher-energy resonance is suppressed relative to the lower-energy resonance, consistent with downhill population redistribution. Importantly, the two detection observables respond differently to this common change in the underlying dynamics: the action-detected features remain substantially broader and display a different redistribution of cross-peak intensity than the corresponding coherent-emission features. This provides a direct spectral manifestation of the observable-dependent projection of population transport.

Further observable-dependent effects emerge when the action-detection weights are made unequal [Fig.~\ref{fig:simulated_amplitude_spectra}(c)]. As expected, changing these weights has little effect on the coherent-emission spectrum because its detection operator is unchanged. By contrast, the action-detected spectrum exhibits pronounced changes in both linewidth and relative spectral weight, including enhanced broadening of the population-derived features. The comparison therefore isolates the role of the final measurement projection: identical microscopic dynamics can produce different experimentally inferred lineshapes when the evolving state is projected onto observables with different state-dependent action weights.

Throughout this Perspective we assume the impulsive excitation limit, in which the antidiagonal width of the rephasing spectrum provides the standard measure of the homogeneous linewidth. Within this limit, Fig.~\ref{fig:antidiagonal_cuts} quantifies the observable dependence of the apparent homogeneous linewidth through normalized antidiagonal cuts of the simulated spectra. The observed linewidth is defined operationally by Eq.~\eqref{eq:Gamma_general_weighted}. In the reference case [Fig.~\ref{fig:antidiagonal_cuts}(a)], both excitonic resonances are systematically broader under action detection than under coherent emitted-field detection, despite the identical Hamiltonian and Liouvillian dynamics used to generate the two signals. The effect is particularly clear because the revised simulations are performed in the pure-dephasing-limited regime, where the direct contribution of excited-state population relaxation to the homogeneous linewidth is negligible. The difference therefore cannot be attributed to the artificially short excited-state lifetime employed in the original illustrative parameter set, but instead reflects the distinct projection of the common nonequilibrium dynamics onto the two detection observables.

Increasing the downhill population-transfer rate \(k_{b\to a}\) [Fig.~\ref{fig:antidiagonal_cuts}(b)] modifies the antidiagonal lineshapes while preserving the systematic distinction between the two detection modalities. Population redistribution contributes to the measured linewidth in both experiments, as represented phenomenologically by Eq.~\eqref{eq:Gamma_general_weighted}, but its spectroscopic manifestation depends on how the resulting nonequilibrium dynamics are projected onto the measured observable. Consequently, the action-detected and coherent-emission linewidths respond differently to the same underlying population-transfer dynamics. The systematic dependence of this observable-specific broadening on the transfer rate is quantified below in Fig.~\ref{fig:linewidth_transport}.

A complementary observable-selective effect appears when the action-detection weights are made unequal [Fig.~\ref{fig:antidiagonal_cuts}(c)]. Because the coherent-emission detection operator is unchanged, its antidiagonal linewidths remain nearly unchanged relative to the reference case. By contrast, the action-detected lineshapes respond directly to the modified state-dependent detection weights and remain substantially broader than their coherent-emission counterparts. Taken together, the three parameter regimes demonstrate that identical microscopic dynamics need not yield identical experimentally inferred homogeneous linewidths: the measured linewidth reflects both the underlying dynamics and the observable through which those dynamics are projected.

\begin{table*}
\caption{\textbf{Observed homogeneous linewidths} extracted from the antidiagonal cuts of Fig.~\ref{fig:antidiagonal_cuts}.}
\begin{ruledtabular}
\begin{tabular}{cccc}
Simulation label (Fig.~\ref{fig:simulated_amplitude_spectra}) &
Mode frequency &
\(\Gamma_{\mathrm{obs}}^{\mathrm{(coh)}}\) (meV) &
\(\Gamma_{\mathrm{obs}}^{\mathrm{(act)}}\) (meV) \\
\hline
(a) Reference & \(\omega_a\) & 5.16 & 8.63 \\
& \(\omega_b\) & 7.10 & 10.99 \\
(b) Enhanced transfer rate & \(\omega_a\) & 5.18 & 8.60 \\
& \(\omega_b\) & 8.91 & 12.68 \\
(c) Unbalanced action detection & \(\omega_a\) & 5.16 & 8.44 \\
& \(\omega_b\) & 7.10 & 10.75 \\
\end{tabular}
\end{ruledtabular}
\label{tab:Simulation_Linewidths}
\end{table*}

We report the values of \(\Gamma_{\mathrm{obs}}\) for both experimental modalities in Table~\ref{tab:Simulation_Linewidths}. Taken together, these values, with Figs.~\ref{fig:simulated_amplitude_spectra} and~\ref{fig:antidiagonal_cuts}, establish that the experimentally extracted homogeneous linewidth is not an intrinsic property of the excitonic system alone, but depends explicitly on the detection operator through which the nonlinear response is projected. The simulations therefore support the central thesis of this Perspective: the experimentally observed dephasing time is an observable-dependent quantity that reflects both the intrinsic quantum dynamics and the measurement channel used to interrogate them.

\begin{figure}[tb]
\centering
\includegraphics[width=8.5cm]{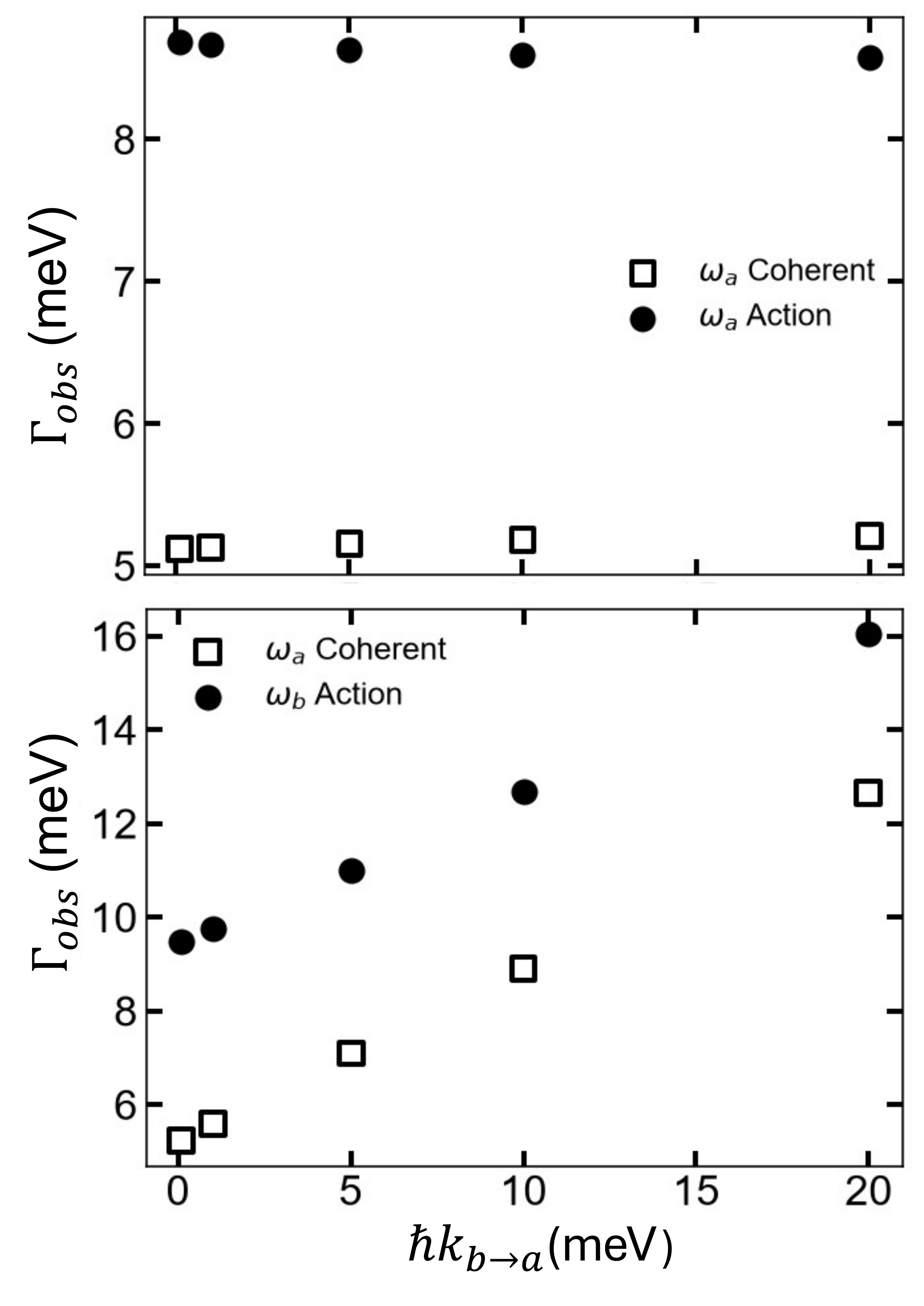}
\caption{
\textbf{Observable-dependent homogeneous linewidth as a function of population transfer.}
Extracted homogeneous linewidth \(\Gamma_{\mathrm{obs}}\) as a function of the population-transfer contribution \(\hbar k_{b\rightarrow a}\) for coherent emitted-field (open squares) and action-detected (filled circles) spectra at the \(\omega_a\) (top) and \(\omega_b\) (bottom) resonances. All microscopic parameters are held fixed as described in Fig.~\ref{fig:simulated_amplitude_spectra}(a) while \(k_{b\rightarrow a}\) is varied, so that the two detection modalities probe the same underlying Hamiltonian and Liouvillian dynamics.
}
\label{fig:linewidth_transport}
\end{figure}

We quantify the dependence of \(\Gamma_{\mathrm{obs}}\) on the downhill population-transfer rate \(k_{b\rightarrow a}\) in Fig.~\ref{fig:linewidth_transport}. At \(\omega_a\) [Fig.~\ref{fig:linewidth_transport}(a)], the linewidths remain nearly independent of population transfer but exhibit a pronounced observable-dependent offset, with action detection yielding the broader feature. At \(\omega_b\) [Fig.~\ref{fig:linewidth_transport}(b)], both linewidths increase with \(k_{b\rightarrow a}\), while their separation grows systematically with increasing population transfer. Thus, the observable dependence itself is resonance-specific: at \(\omega_a\) it appears primarily as a nearly transport-independent linewidth offset, whereas at \(\omega_b\) population redistribution progressively enhances the divergence between the two detection modalities. This behavior demonstrates that population redistribution is projected differently by the two detection observables and provides direct numerical evidence that the homogeneous linewidth inferred from a 2DES spectrum depends jointly on the underlying microscopic dynamics and on the observable through which those dynamics are measured.

The simulations presented here demonstrate that observable-dependent linewidth broadening survives under parameter regimes representative of molecular and polymeric systems. The polymer examples discussed below therefore serve not as validation of the model itself, but as experimental contexts in which these operational distinctions are expected to become relevant.

\section{HJ aggregates in conjugated polymers\label{sec:IV}}
This Perspective is motivated by measurements of homogeneous linewidths in conjugated polymers, carried out in the context of understanding photophysical aggregates in this class of materials, for which we have implemented both experimental schemes discussed here --- see Refs.~\citenum{vella2016ultrafast,li2016probing,gregoire2017excitonic,gutierrez2021frenkel,zheng2024unveiling,kantrow2024quantum}. Before discussing the experimental observations, it is useful to distinguish two conceptually separate aspects of the problem. The first concerns the intrinsic microscopic dynamics governing optical coherence, which are determined by the material Hamiltonian, system--bath interactions, and excitonic structure. In the polymer systems discussed below, these intrinsic dynamics are responsible for the remarkably weak temperature dependence of the homogeneous linewidth, as previously attributed to HJ-aggregate excitonic structure. The second concerns how those same nonequilibrium dynamics are projected into an experimentally recorded signal through a specified detection observable. The present Perspective addresses this latter question. Accordingly, the observable-dependent framework proposed here should be viewed as complementary to, rather than replacing, the microscopic mechanisms responsible for the underlying coherence dynamics.

The examples discussed below are intended to illustrate how the observable-dependent framework provides a unified interpretation of representative experimental observations reported across a variety of material systems. Because these studies were not originally designed as controlled comparisons between coherent-emission and action-detected measurements performed on identical samples under identical conditions, they should not be interpreted as definitive experimental validation of the framework. Rather, they illustrate that the proposed interpretation is consistent with a broad range of reported observations. Across these studies, we observe an apparently anomalous weak temperature dependence of $\Gamma_{\mathrm{obs}}$ over 4--300\,K~\cite{gutierrez2021exciton,kantrow2026understanding} in a progression of solid-state microstructures with increasing chain flexibility, structural order, donor--acceptor character, and aggregate coherence. A systematic comparison of coherent-emission and action-detected measurements on the same family of conjugated polymer systems is presented in Ref.~\citenum{Kantrow2026Anomalous}, providing a direct experimental realization of the observable-dependent framework proposed here. The observed homogeneous linewidth varies from $\sim20$ to $90$\,meV across the series, yet remains nearly temperature independent, within $\sim10\%$, for each material. This behavior is striking because strongly vibronically active systems are generally expected to exhibit pronounced thermal pure dephasing. Representative values are summarized in Table~\ref{tab:linewidth_comparison}, highlighting the systematic difference between coherent emitted-field and action-detected measurements.

\begin{table*}[t]
\centering
\caption{\textbf{Representative homogeneous linewidths extracted from coherent emitted-field and action-detected multidimensional spectroscopy in conjugated polymer systems.}
Values are reported for comparable temperature ranges; see cited references for full experimental details.}
\begin{ruledtabular}
\begin{tabular}{lcccc}
Material & Detection modality & $\Gamma_{\mathrm{obs}}$ (meV) & Temperature range & Ref. \\
\hline
PBTTT & Coherent emission & $46 \pm 2$ & 4--300 K & \citenum{gutierrez2021exciton,kantrow2026understanding,Kantrow2026Anomalous} \\
PBTTT & Action (PL-detected) & $75 \pm 10$ & 4--300 K & \citenum{gutierrez2021exciton,kantrow2026understanding,gutierrez2021frenkel,Kantrow2026Anomalous} \\
P3HT  & Coherent emission & $\sim 40$ & 4--300 K & \citenum{kantrow2026understanding} \\
P3HT  & Action (PL-detected) & $\sim 90$ & 8 K & \citenum{gregoire2017excitonic,Kantrow2026Anomalous} \\
\end{tabular}
\end{ruledtabular}
\label{tab:linewidth_comparison}
\end{table*}

We rationalized this behavior within an exciton--vibrational framework as a consequence of the hybrid HJ aggregate electronic structure characteristic of these polymers~\cite{spano2014h}. Here, ``HJ'' denotes the coexistence of J-like intrachain excitonic delocalization and H-like interchain Coulomb coupling, producing an excitonic manifold that supports both bright band-edge states and nearby dark relaxation channels. The key point is that exciton--vibrational coupling alone does not determine the quantum dynamics; rather, the aggregate band structure determines which sectors of the environmental fluctuation spectrum are sampled by the optically prepared excitation. To test this interpretation, we implemented a triangulation strategy mirroring the central thesis of this Perspective: selected materials were interrogated using both detection modalities together with resonance Raman intensity analysis as an independent probe of exciton--vibrational coupling~\cite{gutierrez2021exciton,kantrow2026understanding}. Systematically, we find that 
$\Gamma_{\mathrm{obs}}^{(\mathrm{coh})}
<
\Gamma_{\mathrm{obs}}^{(\mathrm{inc})}$,
supporting the conclusion that the inferred linewidth depends on the observable through which the nonequilibrium dynamics are projected. The nearly temperature-independent intrinsic coherence dynamics have previously been attributed to the HJ-aggregate excitonic structure. Within the present framework, these intrinsic dynamics constitute the common underlying evolution that is subsequently projected onto different detection observables, leading to observable-dependent experimentally inferred linewidths.

The semiconductor polymer PBTTT provides an especially clear example. This polymer forms liquid-crystalline-like mesophases associated with its ribbon-like backbone, promoted by the rigid thienothiophene unit and side-chain interdigitation. We measure
$\Gamma_{\mathrm{obs}}^{(\mathrm{coh})}\sim46\pm2$\,meV and
$\Gamma_{\mathrm{obs}}^{(\mathrm{inc})}\sim75\pm10$\,meV
(Table~\ref{tab:linewidth_comparison}), with both values remaining nearly unchanged over the full 4--300\,K range~\cite{gutierrez2021exciton,kantrow2026understanding}. The films used in both measurements were prepared from the same polymer batch under closely matched processing conditions, supporting that the observed difference is intrinsic to the detection modality.

A similarly instructive contrast is provided by P3HT (Table~\ref{tab:linewidth_comparison}), the canonical semicrystalline conjugated polymer, whose high-molecular-weight films ($M_W \gtrsim 50$\,kg/mol) adopt a two-phase morphology consisting of crystalline domains embedded in amorphous chain-entangled regions~\cite{reid2012influence,paquin2013two}. Contrary to a common simplification in the literature, the high crystallinity of P3HT does not imply a narrow electronic density of states. In PL-detected multidimensional measurements, we reported $\Gamma_{\mathrm{obs}}^{(\mathrm{inc})}\sim90$\,meV at 8\,K~\cite{gregoire2017excitonic}, whereas coherent emitted-field measurements yield $\Gamma_{\mathrm{obs}}^{(\mathrm{coh})}\sim40$\,meV over the full 4--300\,K range~\cite{kantrow2026understanding}. The persistence of this factor-of-two discrepancy across two chemically and morphologically distinct polymer systems strongly supports the detection-dependent linewidth inequality.

These two case studies provide direct experimental support for the central claim of this Perspective. The persistence of the linewidth inequality across heterogeneous HJ-aggregate microstructures identifies a regime in which pulse ordering and delayed population conversion can mimic additional dephasing and obscure the interpretation of $T_{2,\mathrm{eff}}$, motivating the discussion of pseudo-dephasing and its experimental diagnostics in the next section. The dedicated same-material comparison reported in Ref.~\citenum{Kantrow2026Anomalous} provides an experimental bridge between the conceptual framework developed in this Perspective and quantitative observable-dependent spectroscopy of complex molecular systems. Future experiments combining coherent-emission and action-detected multidimensional spectroscopy on the same sample under identical experimental conditions will provide the most direct quantitative tests of the observable-dependent framework developed here.

\section{Incoherent population mixing and pseudo-dephasing\label{sec:V}}
The polymer case studies above point to a broader interpretive
challenge: in population-detected multidimensional spectroscopy,
linewidth broadening does not necessarily reflect loss of optical
phase coherence. Delayed population dynamics can instead generate
spectral structure that mimics dephasing.

The clearest manifestation of this effect arises from incoherent
population mixing. In this regime, populations generated by
different pulse interactions evolve independently and subsequently
interact through nonlinear kinetics during or after the pulse
sequence. The measured signal therefore contains contributions that
do not originate from a single coherent Liouville pathway, but from
population-level mixing. Consequently, cross peaks, linewidths, and
waiting-time dynamics can acquire contributions that are not directly
related to optical coherence decay.

This behavior has been documented explicitly in our earlier work on
phase-modulated 2DES, where incoherent population mixing was shown to
generate cross peaks and apparent spectral broadening without
corresponding additional coherent pathway
content~\cite{gregoire2017incoherent,bargigia2022identifying}. In
semiconducting polymer films, processes such as exciton migration,
exciton–exciton annihilation, and downhill relaxation into
lower-energy emissive states can continue on timescales comparable
to, or longer than, the inter-pulse delays~\cite{gregoire2017incoherent}.
The detected population can therefore continue to evolve after the
coherent light–matter interactions have occurred, so that the
time-integrated action signal reflects both the coherent response
encoded by the pulse sequence and its subsequent population
processing.

When this additional spectral structure can be represented
phenomenologically as linewidth broadening, one may write
\begin{equation}
\Gamma_{\mathrm{obs}}^{\prime}
=
\Gamma_{\mathrm{obs}}^{(\mathrm{act})}
+
\Gamma_{\mathrm{kin}}.
\end{equation}
where $\Gamma_{\mathrm{obs}}^{(\mathrm{act})}$ denotes the
observable-dependent optical decoherence described by
equation~\eqref{eq:Gammaobs}, while $\Gamma_{\mathrm{kin}}$ represents
an effective broadening associated with nonlinear population kinetics
entering the time-integrated observable of
equation~\eqref{eq:Sact_integrated}. This decomposition is
phenomenological: incoherent population mixing can also generate new
spectral features rather than simple broadening. In either case,
$\Gamma_{\mathrm{kin}}$ should not be interpreted as an additional
intrinsic contribution to the material coherence decay rate.

A key practical question is therefore how to determine when such
incoherent contributions are present. A particularly direct
diagnostic arises from cross-talk between modulation channels in the
four-pulse sequence, as demonstrated in
ref.~\citenum{bargigia2022identifying}. In photocurrent-detected
two-dimensional spectroscopy, incoherent population mixing produces
measurable correlations between signals associated with nominally
independent coherence-delay variables, providing a fingerprint of
nonlinear population mixing contributions to the measured
lineshape.

Such effects motivate the useful distinction between
observable-dependent dephasing and \emph{pseudo-dephasing}. The
former describes genuine loss of optical coherence as projected
through a specified detection observable; the latter denotes apparent
spectral broadening or related lineshape changes generated by
population dynamics after the relevant optical coherence has been
created. Recognizing~\cite{bargigia2022identifying,gutierrez2021frenkel}
and mitigating~\cite{faitz2024spectrometer} this regime is therefore
essential for interpreting action-detected 2DES of heterogeneous
materials, where population redistribution can occur throughout the
detection window represented by
equation~\eqref{eq:Sact_integrated} and is intrinsic to the
photoexcited-state landscape.

\section{Perspective: designing observables to probe materials function\label{sec:VI}}

The central message of this Perspective is that the detection observable must be treated as a fundamental part of the 2DES experiment. The apparent homogeneous linewidth and inferred dephasing time do not arise solely from the material Hamiltonian and its bath-induced fluctuations, but from how the nonequilibrium density matrix—the photoexcited state—is projected onto an experimentally accessible observable defined by the detection scheme. The choice of detection modality therefore directly shapes which aspects of the underlying dynamics and materials function are experimentally accessible.

This recognition suggests a broader conceptual shift for multidimensional spectroscopy. Historically, the field has focused primarily on pulse-sequence design, phase matching, and pathway selection as the principal means of isolating coherent dynamics. The framework developed here shows that the detection operator is equally fundamental. The opportunity is therefore not only to engineer excitation pathways, but also to engineer observables that are maximally sensitive to the functional degrees of freedom most relevant to the material under study.

In semiconducting polymers, such observable engineering may emphasize exciton migration, multiexciton interactions, or relaxation into dark-state manifolds associated with hybrid HJ aggregate structure~\cite{gutierrez2021exciton,kantrow2026understanding}; in polymer-based solar cells, it may distinguish carrier-generation and charge-recombination pathways~\cite{vella2016ultrafast}; and in organic and hybrid organic–inorganic semiconductors, it may enhance sensitivity to biexciton correlations and binding~\cite{thouin2018stable,gutierrez2021frenkel,zheng2024unveiling,koch2025spectroscopic,koch2026biexcitons}. In this broader view, observable engineering transforms 2DES from a general probe of coherence into a targeted probe of structure–dynamics–function relationships.

A particularly promising extension is toward the identification and quantification of many-body correlations in quantum materials. In strongly correlated systems, including candidate quantum spin liquids, excitonic insulators, and low-dimensional magnets, the relevant degrees of freedom cannot generally be reduced to independent single-particle excitations. Multidimensional spectroscopies can encode higher-order correlations in cross peaks, spectral correlations, and nonlinear response pathways, but the extent to which these features reveal specific many-body dynamics depends critically on the observable through which the response is detected. By comparing or designing coherent-emission, action-detected, and hybrid measurement schemes, it may become possible to enhance sensitivity to selected correlated sectors of the dynamics while suppressing competing incoherent contributions. Developing operational mappings between multidimensional spectra and quantitative measures of many-body correlations—and, ultimately, experimentally accessible witnesses of quantum correlations—represents an important direction for connecting coherent spectroscopy with quantum-materials discovery.

More generally, the concept of observability developed here is not restricted to coherent spectroscopy. It raises the broader question of which dynamical properties of an open quantum system remain experimentally accessible under a specified measurement protocol. A nonlinear spectroscopic experiment does not measure microscopic dynamics directly; it measures a projection of those dynamics onto a chosen observable. Consequently, experimentally inferred quantities such as dephasing rates, population-transfer contributions, and effective linewidths should be understood operationally rather than assumed to be intrinsic material constants independent of measurement.

The analytical framework and numerical simulations considered here show that the divergence between detection observables persists across experimentally relevant parameter regimes. Observable-dependent linewidths are therefore not a peculiarity of a specific model, but a general consequence of projecting nonequilibrium dynamics onto different measured quantities. This distinction extends beyond the specific comparison between coherent field-emission and action-detected multidimensional spectroscopies considered here.

We therefore propose \emph{observable engineering} as a general principle for multidimensional spectroscopy: experimentally accessible information is determined jointly by the system dynamics and by the observable through which those dynamics are projected. The detection observable is not simply the final stage of the experiment; it helps define the physical meaning of the measured spectrum and determines which aspects of the underlying dynamics become experimentally accessible. Treating it as a first-class design variable provides a framework for interpreting existing measurement schemes, designing new ones, and establishing more rigorous structure–dynamics–function relationships in complex molecular and condensed-matter systems.

This perspective also leads naturally from observability to \emph{identifiability}. Once the measurement projection has been specified, the next question is not merely whether a dynamical process contributes to the signal, but whether its underlying physical parameters can be uniquely inferred from the available observables. Determining which parameters remain identifiable under different measurement projections provides a natural next step toward a more general theory of information extraction from nonlinear spectroscopy.

\section*{Supplementary Material}
The Supplementary Material file contains section SI: "OBSERVABLE-WEIGHTED POPULATION EXCHANGE AND FINITE-TEMPERATURE LIMITS", containing derivation of equations~\eqref{eq:low-temp} and \eqref{eq:high-temp}, and section SII: "2DES LINESHAPE AS A FUNCTION OF EXCITED-STATE TRANSPORT", preenting spectra used to construct Fig.~\ref{fig:linewidth_transport}.

\begin{acknowledgments}
CSA and ERB dedicate this Perspective to the memory of Mark Ratner, a role-model, a mentor, and a friend. 

HL and CSA acknowledge funding from the Government of Canada (Canada Excellence Research Chair CERC-2022-00055). CSA acknowledges support from the Institut Courtois, Facult\'e des arts et des sciences, Universit\'e de Montr\'eal (Chaire de recherche de direction de l'Institut Courtois) and from the Natural Science and Engineering Research Council of Canada (NSERC Discovery Grant RGPIN-2024-05893). 
CSA and SPO gratefully acknowledge support from the training program \emph{CREATE for Accelerated Discovery} (AccelD) hosted by the Acceleration Consortium and co-led by the Institut Courtois (NSERC Grant \#596133-2025). 
ERB acknowledges funding from the National Science Foundation (CHE-2404788), Robert A.\ Welch Foundation (E-1337), the Department of Energy supported this research through Award No. 11937-PO147716. ERB gratefully acknowledges funding from IVADO for a Visiting Professorship at the Institut Courtois, Universit\'e de Montr\'eal. 
\end{acknowledgments}

\section*{Author declarations}

\subsection*{Conflict of Interest}
The authors have no conflicts to disclose.

\subsection*{Author Contributions}
SPO: Investigation (numerical simulations), Software; HL: Supervision, Conceptualization; ERB: Conceptualization; CSA: Conceptualization (lead), Supervision, Project administration, Writing—original draft; All authors: Writing—review \& editing.

\subsection*{Use of Generative Artificial Intelligence}
In compliance with institutional guidelines of the Universit\'e de Montr\'eal, generative artificial intelligence tools were used to assist with the editing of language and stylistic refinement of parts of the manuscript and to assist in the synthesis of the literature. These tools were not used to generate scientific content, perform analysis, or influence the interpretation of results. All content has been reviewed and validated by the authors, who assume full responsibility for the manuscript.

\section*{Data availability}
In accordance to the Data Management Plan of the Canada Excellence Research Chair in Light-Matter interactions (\url{http://hdl.handle.net/1866/33427}), the numerical data and code that support the findings of this study are openly available in the Borealis Dataverse Repository at \url{https://doi.org/10.5683/SP4/UKSLI2}.

%

\clearpage
\clearpage
\newpage
\setcounter{page}{1}
\setcounter{section}{0}
\setcounter{subsection}{0}
\setcounter{figure}{0} 
\setcounter{equation}{0}
\setcounter{table}{0}
\renewcommand{\thefigure}{S\arabic{figure}} 
\renewcommand{\thetable}{S\Roman{table}}
\renewcommand{\thepage}{S\arabic{page}} 
\renewcommand{\theequation}{S\arabic{equation}} 
\renewcommand{\thesection}{S\Roman{section}}
\renewcommand{\thesubsection}{\thesection.\Alph{subsection}}
\renewcommand{\thesubsubsection}{\thesubsection.\arabic{subsubsection}}

\onecolumngrid

\noindent
{\large \textbf{Supplemental Material for \textit{``Detection Defines Dephasing in Two-Dimensional Electronic Spectroscopy of Materials: Coherent Field Emission versus Incoherent Population Observables''}, Paiva-Ortega~\textit{et al.}}}

\section{Observable-weighted population exchange and finite-temperature limits}
\label{sec:SI_population_exchange}

In the main text, the experimentally inferred linewidth is interpreted as an
observable-dependent quantity obtained by projecting a common nonequilibrium
dynamics onto a specified detection operator. Here we derive this result for
population exchange between two excited states. The derivation is formulated
first for an arbitrary initial population distribution and then specialized to
the optically prepared pure-state initial condition used in the main text.

Throughout this section, the rate constants \(k\) have units of inverse time.
The corresponding contribution to an energy linewidth is obtained by
multiplication by \(\hbar\). Equivalently, \(\hbar\) may be absorbed into the
definition of the rates when they are reported in energy units.

\subsection{Population dynamics and detection observable}

Consider two excited states \(a\) and \(b\), with populations collected into
the vector
\begin{equation}
    \mathbf{p}(t)
    =
    \begin{pmatrix}
        p_a(t)\\
        p_b(t)
    \end{pmatrix}.
    \label{eq:SI_population_vector}
\end{equation}
The populations evolve according to the kinetic equation
\begin{equation}
    \frac{d\mathbf{p}}{dt}
    =
    \mathbf{K}\mathbf{p},
    \label{eq:SI_rate_equation}
\end{equation}
with rate matrix
\begin{equation}
    \mathbf{K}
    =
    \begin{pmatrix}
        -k_a-k_{a\rightarrow b} & k_{b\rightarrow a}\\[4pt]
        k_{a\rightarrow b} & -k_b-k_{b\rightarrow a}
    \end{pmatrix}.
    \label{eq:SI_rate_matrix}
\end{equation}
Here, \(k_a\) and \(k_b\) denote irreversible population-loss rates from
states \(a\) and \(b\), respectively, while \(k_{a\rightarrow b}\) and
\(k_{b\rightarrow a}\) describe bidirectional population exchange within the
excited-state manifold.

For an action-detected measurement, the signal is a weighted projection of
the population vector,
\begin{equation}
    S(t)
    =
    \boldsymbol{\eta}^{\mathsf T}\mathbf{p}(t)
    =
    \eta_a p_a(t)+\eta_b p_b(t),
    \label{eq:SI_detected_signal}
\end{equation}
where
\begin{equation}
    \boldsymbol{\eta}
    =
    \begin{pmatrix}
        \eta_a\\
        \eta_b
    \end{pmatrix}
\end{equation}
contains the state-dependent action yields and detection efficiencies. The
formal solution of Eq.~\eqref{eq:SI_rate_equation} is
\begin{equation}
    \mathbf{p}(t)
    =
    e^{\mathbf{K}t}\mathbf{p}(0),
    \label{eq:SI_formal_solution}
\end{equation}
so that
\begin{equation}
    S(t)
    =
    \boldsymbol{\eta}^{\mathsf T}
    e^{\mathbf{K}t}
    \mathbf{p}(0).
    \label{eq:SI_formal_signal}
\end{equation}

\subsection{General observable-dependent logarithmic decay rate}

We define the instantaneous effective decay rate of the measured action
signal by
\begin{equation}
    \gamma_{\mathrm{act}}(t)
    \equiv
    -
    \frac{d}{dt}\ln S(t).
    \label{eq:SI_instantaneous_rate_definition}
\end{equation}
Using Eqs.~\eqref{eq:SI_rate_equation} and
\eqref{eq:SI_detected_signal}, this may be written in the general form
\begin{equation}
    \gamma_{\mathrm{act}}(t)
    =
    -
    \frac{
        \boldsymbol{\eta}^{\mathsf T}
        \mathbf{K}
        e^{\mathbf{K}t}
        \mathbf{p}(0)
    }{
        \boldsymbol{\eta}^{\mathsf T}
        e^{\mathbf{K}t}
        \mathbf{p}(0)
    }.
    \label{eq:SI_general_time_dependent_rate}
\end{equation}
At the initial time,
\begin{equation}
    \gamma_{\mathrm{act}}(0)
    =
    -
    \frac{
        \boldsymbol{\eta}^{\mathsf T}
        \mathbf{K}\mathbf{p}(0)
    }{
        \boldsymbol{\eta}^{\mathsf T}\mathbf{p}(0)
    }.
    \label{eq:SI_general_initial_rate}
\end{equation}
Equation~\eqref{eq:SI_general_initial_rate} is the general
observable-projection result for the initial logarithmic slope. It shows
explicitly that the measured decay rate depends jointly on the population
dynamics, the initial state, and the detection observable.

For an arbitrary initial condition
\begin{equation}
    \mathbf{p}(0)
    =
    \begin{pmatrix}
        p_a^0\\
        p_b^0
    \end{pmatrix},
    \qquad
    p_a^0+p_b^0=1,
    \label{eq:SI_general_initial_condition}
\end{equation}
Eq.~\eqref{eq:SI_general_initial_rate} becomes
\begin{align}
    \gamma_{\mathrm{act}}(0)
    &=
    \frac{
        \eta_a k_a p_a^0
        +
        \eta_b k_b p_b^0
    }{
        \eta_a p_a^0+\eta_b p_b^0
    }
    \nonumber\\[4pt]
    &\quad+
    \frac{
        (\eta_a-\eta_b)
        \left(
            k_{a\rightarrow b}p_a^0
            -
            k_{b\rightarrow a}p_b^0
        \right)
    }{
        \eta_a p_a^0+\eta_b p_b^0
    }.
    \label{eq:SI_general_initial_rate_expanded}
\end{align}
The first term is the detection-weighted contribution from irreversible
population loss. The second is the contribution from population exchange.
It depends on both the net population flux and the difference between the
detection weights.

Several consequences follow directly from
Eq.~\eqref{eq:SI_general_initial_rate_expanded}. First, exchange does not
contribute when
\begin{equation}
    \eta_a=\eta_b,
    \label{eq:SI_balanced_detection}
\end{equation}
because a balanced observable is insensitive to redistribution between the
two states. Second, exchange does not contribute to the initial slope when
the initial populations satisfy
\begin{equation}
    k_{a\rightarrow b}p_a^0
    =
    k_{b\rightarrow a}p_b^0,
    \label{eq:SI_zero_net_flux}
\end{equation}
because the exchange dynamics initially carry no net population flux.

\subsection{Optical preparation of state $b$}

The simulations and main-text analysis consider selective optical
preparation of the higher-energy state \(b\),
\begin{equation}
    \mathbf{p}(0)
    =
    \begin{pmatrix}
        0\\
        1
    \end{pmatrix}.
    \label{eq:SI_pure_b_initial_condition}
\end{equation}
For this initial condition,
\begin{equation}
    \dot{\mathbf{p}}(0)
    =
    \mathbf{K}\mathbf{p}(0)
    =
    \begin{pmatrix}
        k_{b\rightarrow a}\\[4pt]
        -k_b-k_{b\rightarrow a}
    \end{pmatrix}.
    \label{eq:SI_pure_b_initial_derivative}
\end{equation}
The uphill rate \(k_{a\rightarrow b}\) does not enter the initial derivative
because \(p_a(0)=0\). The initial logarithmic decay rate is therefore
\begin{equation}
    \gamma_{\mathrm{act}}^{(b)}(0)
    =
    k_b
    +
    k_{b\rightarrow a}
    \left(
        1-\frac{\eta_a}{\eta_b}
    \right).
    \label{eq:SI_pure_b_action_rate}
\end{equation}
The observable-dependent exchange contribution is
\begin{equation}
    \Delta\gamma_{\mathrm{ex}}^{(b)}
    =
    k_{b\rightarrow a}
    \left(
        1-\frac{\eta_a}{\eta_b}
    \right),
    \label{eq:SI_pure_b_exchange_rate}
\end{equation}
or, expressed as an energy linewidth,
\begin{equation}
    \boxed{
    \Delta\Gamma_{\mathrm{ex}}^{(b)}
    =
    \hbar k_{b\rightarrow a}
    \left(
        1-\frac{\eta_a}{\eta_b}
    \right)
    }.
    \label{eq:SI_pure_b_exchange_linewidth}
\end{equation}
This result is the microscopic realization of the observable-weighted
population-redistribution contribution introduced phenomenologically in the
main text.

If optical dephasing and population relaxation are included, the total
observed linewidth associated with resonance \(b\) may be written as
\begin{equation}
    \Gamma_{\mathrm{obs}}^{(b)}
    =
    \Gamma_{\phi}^{(b)}
    +
    \Gamma_{1}^{(b)}
    +
    \hbar k_{b\rightarrow a}
    \left(
        1-\frac{\eta_a}{\eta_b}
    \right),
    \label{eq:SI_total_pure_b_linewidth}
\end{equation}
within the approximately exponential and separable regime discussed in the
main text.

The final term in Eq.~\eqref{eq:SI_total_pure_b_linewidth} is the microscopic realization of the
observable-weighted population-redistribution contribution
\(w_{\mathrm{trans}}^{(\alpha)}\Gamma_{\mathrm{trans}}\) introduced in the
main text. The state-dependent action yields and subsequent population
evolution enter through the measurement projection and therefore do not
define an additional independent linewidth contribution.

\subsection{Finite-temperature parametrization}

At finite temperature, both downhill and uphill exchange rates may be
nonzero. Detailed balance gives
\begin{equation}
    \frac{k_{b\rightarrow a}}{k_{a\rightarrow b}}
    =
    \exp\left[
        \frac{\hbar(\omega_b-\omega_a)}{k_{\mathrm B}T}
    \right],
    \qquad
    \omega_b>\omega_a.
    \label{eq:SI_detailed_balance}
\end{equation}
We define the total exchange rate
\begin{equation}
    k_{\mathrm{ex}}
    \equiv
    k_{b\rightarrow a}
    +
    k_{a\rightarrow b},
    \label{eq:SI_total_exchange_rate}
\end{equation}
and the thermodynamic asymmetry
\begin{equation}
    \chi(T)
    \equiv
    \frac{
        k_{b\rightarrow a}
        -
        k_{a\rightarrow b}
    }{
        k_{b\rightarrow a}
        +
        k_{a\rightarrow b}
    }.
    \label{eq:SI_asymmetry_definition}
\end{equation}
Using Eq.~\eqref{eq:SI_detailed_balance},
\begin{equation}
    \chi(T)
    =
    \tanh\left[
        \frac{\hbar(\omega_b-\omega_a)}
        {2k_{\mathrm B}T}
    \right].
    \label{eq:SI_asymmetry_temperature}
\end{equation}
The individual rates are therefore
\begin{align}
    k_{b\rightarrow a}
    &=
    \frac{k_{\mathrm{ex}}}{2}
    \left[1+\chi(T)\right],
    \label{eq:SI_downhill_rate_parametrization}\\[4pt]
    k_{a\rightarrow b}
    &=
    \frac{k_{\mathrm{ex}}}{2}
    \left[1-\chi(T)\right].
    \label{eq:SI_uphill_rate_parametrization}
\end{align}

For the pure-\(b\) initial condition of
Eq.~\eqref{eq:SI_pure_b_initial_condition}, only the downhill rate enters
the initial logarithmic slope. Substitution of
Eq.~\eqref{eq:SI_downhill_rate_parametrization} into
Eq.~\eqref{eq:SI_pure_b_exchange_linewidth} gives
\begin{equation}
    \boxed{
    \Delta\Gamma_{\mathrm{ex}}^{(b)}(T)
    =
    \frac{\hbar k_{\mathrm{ex}}}{2}
    \left[1+\chi(T)\right]
    \left(
        1-\frac{\eta_a}{\eta_b}
    \right)
    }.
    \label{eq:SI_finite_temperature_pure_b_linewidth}
\end{equation}
Equation~\eqref{eq:SI_finite_temperature_pure_b_linewidth} is a
finite-temperature parametrization of the initial downhill-transfer
contribution. It should not be interpreted as an independent sum of uphill
and downhill contributions to the initial slope. The uphill process enters
the subsequent finite-time dynamics only after population has accumulated
in state \(a\).

\subsection{Limiting cases}

Equation~\eqref{eq:SI_finite_temperature_pure_b_linewidth} satisfies the
relevant limiting cases.

\paragraph{Low-temperature limit.}
When
\begin{equation}
    k_{\mathrm B}T
    \ll
    \hbar(\omega_b-\omega_a),
\end{equation}
one has
\begin{equation}
    \chi(T)\rightarrow 1,
    \qquad
    k_{a\rightarrow b}\ll k_{b\rightarrow a},
\end{equation}
and therefore
\begin{equation}
    \Delta\Gamma_{\mathrm{ex}}^{(b)}
    \rightarrow
    \hbar k_{b\rightarrow a}
    \left(
        1-\frac{\eta_a}{\eta_b}
    \right).
    \label{eq:SI_low_temperature_limit}
\end{equation}

\paragraph{High-temperature limit.}
When
\begin{equation}
    k_{\mathrm B}T
    \gg
    \hbar(\omega_b-\omega_a),
\end{equation}
one has
\begin{equation}
    \chi(T)\rightarrow 0,
    \qquad
    k_{b\rightarrow a}
    \simeq
    k_{a\rightarrow b}
    \simeq
    \frac{k_{\mathrm{ex}}}{2},
\end{equation}
so that
\begin{equation}
    \Delta\Gamma_{\mathrm{ex}}^{(b)}
    \rightarrow
    \frac{\hbar k_{\mathrm{ex}}}{2}
    \left(
        1-\frac{\eta_a}{\eta_b}
    \right).
    \label{eq:SI_high_temperature_limit}
\end{equation}
Although the uphill and downhill rates are equal in this limit, the initial
slope following pure-\(b\) preparation contains only the instantaneous
downhill flux out of state \(b\).

\paragraph{Balanced detection.}
For
\begin{equation}
    \eta_a=\eta_b,
\end{equation}
the exchange contribution vanishes at every temperature,
\begin{equation}
    \Delta\Gamma_{\mathrm{ex}}^{(b)}(T)=0.
    \label{eq:SI_balanced_detection_limit}
\end{equation}
Population redistribution remains present dynamically, but it is invisible
to an observable that assigns equal weight to the two participating states.

\subsection{Thermal and mixed initial populations}

A genuinely bidirectional contribution to the initial logarithmic slope
requires an initial population distribution with both
\(p_a^0\neq 0\) and \(p_b^0\neq 0\). For such a mixed nonequilibrium
initial state, Eq.~\eqref{eq:SI_general_initial_rate_expanded} gives
\begin{equation}
    \Delta\gamma_{\mathrm{ex}}(0)
    =
    \frac{
        (\eta_a-\eta_b)
        \left(
            k_{a\rightarrow b}p_a^0
            -
            k_{b\rightarrow a}p_b^0
        \right)
    }{
        \eta_a p_a^0+\eta_b p_b^0
    }.
    \label{eq:SI_mixed_state_exchange_rate}
\end{equation}
Both uphill and downhill rates now enter explicitly.

For a thermal distribution within the two-state excited manifold,
\begin{equation}
    p_a^{\mathrm{eq}}
    =
    \frac{k_{b\rightarrow a}}{k_{\mathrm{ex}}},
    \qquad
    p_b^{\mathrm{eq}}
    =
    \frac{k_{a\rightarrow b}}{k_{\mathrm{ex}}},
    \label{eq:SI_equilibrium_populations}
\end{equation}
detailed balance implies
\begin{equation}
    k_{a\rightarrow b}p_a^{\mathrm{eq}}
    =
    k_{b\rightarrow a}p_b^{\mathrm{eq}}.
    \label{eq:SI_equilibrium_flux_balance}
\end{equation}
Consequently,
\begin{equation}
    \Delta\gamma_{\mathrm{ex}}^{\mathrm{eq}}(0)=0.
    \label{eq:SI_equilibrium_exchange_slope}
\end{equation}
At thermal equilibrium there is no net exchange flux, so population exchange
does not contribute to the initial logarithmic decay even when the detection
weights are unequal. Observable-dependent exchange broadening therefore
requires a nonequilibrium population distribution, as is naturally produced
by optical excitation.

\subsection{Finite-time bidirectional dynamics}

Even when the system is initially prepared entirely in state \(b\), both
exchange rates influence the measured signal at finite times. Once downhill
transfer produces a nonzero population \(p_a(t)\), the uphill channel
\(a\rightarrow b\) enters the evolution through the full propagator
\(e^{\mathbf{K}t}\). The corresponding instantaneous effective decay rate is
\begin{equation}
    \gamma_{\mathrm{act}}^{(b)}(t)
    =
    -
    \frac{
        \boldsymbol{\eta}^{\mathsf T}
        \mathbf{K}
        e^{\mathbf{K}t}
        \begin{pmatrix}
            0\\
            1
        \end{pmatrix}
    }{
        \boldsymbol{\eta}^{\mathsf T}
        e^{\mathbf{K}t}
        \begin{pmatrix}
            0\\
            1
        \end{pmatrix}
    }.
    \label{eq:SI_finite_time_pure_b_rate}
\end{equation}
Unlike the initial slope,
Eq.~\eqref{eq:SI_finite_time_pure_b_rate} generally depends on both
\(k_{b\rightarrow a}\) and \(k_{a\rightarrow b}\). A linewidth inferred
from a finite temporal window may therefore encode the full bidirectional
exchange dynamics and need not coincide with the initial-slope result.

\subsection{Physical interpretation}

The derivation establishes three distinct points.
First, the contribution of population exchange to an action-detected
linewidth is not determined by the exchange kinetics alone. It is weighted
by the detection observable through the ratio \(\eta_a/\eta_b\). 
Accordingly, within the minimal model considered here, the observed
linewidth contains three distinct contributions: pure dephasing,
population relaxation, and observable-weighted population redistribution.
The action-detection dependence resides in the projection of the
redistribution dynamics onto the chosen observable rather than in a
separate additive decay channel.
Second, following selective optical preparation of state \(b\), the initial
logarithmic slope contains only the instantaneous downhill rate
\(k_{b\rightarrow a}\). Finite temperature changes this rate through detailed
balance, but the uphill rate does not enter independently at \(t=0\).
Third, both uphill and downhill rates enter either for a mixed
nonequilibrium initial population or through the finite-time evolution of
the action signal. These results distinguish the initial-slope linewidth
used in the minimal analytic treatment from the more general
bidirectional dynamics contained in the full Liouvillian propagation.

\section{2DES lineshape as a function of excited-state transport}

\begin{figure*}[ht]
\centering
\includegraphics[width=\linewidth]{Figures/SI_Spectra_EST_CE.pdf}

\caption{
\textbf{Simulated 2DES normalized amplitude rephasing spectra} used to extract the homogeneous linewidth \(\Gamma_{\mathrm{obs}}\) for coherent emission in Fig.~\ref{fig:linewidth_transport}.
}
\label{fig:simulated_amplitude_spectra-CE}

\vspace{0.5em}

\begin{minipage}{0.96\linewidth}
\noindent\textbf{Alt text:}
Five simulated normalized rephasing 2DES amplitude spectra for coherent emitted-field detection, shown for increasing downhill population-transfer rates from 0.1 to 20\,meV. Each spectrum contains two diagonal resonances and associated cross-peak features. As the transfer rate increases, the higher-energy resonance and cross-peak structure change progressively in intensity and lineshape. These spectra are used to extract the coherent-emission homogeneous linewidths reported in Fig.~4.
\end{minipage}

\end{figure*}

\begin{figure*}[ht]
\centering
\includegraphics[width=\linewidth]{Figures/SI_Spectra_EST_AD.pdf}

\caption{
\textbf{Simulated 2DES normalized amplitude rephasing spectra} used to extract the homogeneous linewidth \(\Gamma_{\mathrm{obs}}\) for action detection in Fig.~\ref{fig:linewidth_transport}.
}
\label{fig:simulated_amplitude_spectra-AD}

\vspace{0.5em}

\begin{minipage}{0.96\linewidth}
\noindent\textbf{Alt text:}
Five simulated normalized rephasing 2DES amplitude spectra for action detection, shown for increasing downhill population-transfer rates from 0.1 to 20\,meV. The spectra contain two diagonal resonances and associated cross-peak features, with broader lineshapes than the corresponding coherent-emission spectra. Increasing population transfer progressively modifies the relative intensities and lineshapes of the resonances and cross peaks. These spectra are used to extract the action-detected homogeneous linewidths reported in Fig.~4.
\end{minipage}

\end{figure*}

\end{document}